\documentclass[twocolumn,preprintnumbers,amsmath,amssymb]{revtex4}

\usepackage{graphicx}
\usepackage{dcolumn}
\usepackage{bm}
\usepackage{rotating}

\usepackage{bbm}
\usepackage{verbatim}
\usepackage{afterpage}

\usepackage{color}
\definecolor{light blue}{rgb}{0.2,0.2,0.7}
\definecolor{darkblue}{rgb}{0,0.25,0.5}
\definecolor{redbrown}{rgb}{0.875,0.25,0.125}
\definecolor{darkgreen}{rgb}{0,0.5,0}

\newcommand{\bra}[1]{\ensuremath{\langle #1 \vert}}
\newcommand{\ket}[1]{\ensuremath{\vert #1  \rangle}}

\renewcommand{\b}[1]{\ensuremath{\mathbf{#1}}}
\renewcommand{\H}{\ensuremath{\text{H}}}
\renewcommand{\l}{\ensuremath{\lambda}}

\newcommand{\GGA}{\ensuremath{\text{GGA}}}

\newcommand{\n}{\ensuremath{\nabla}}

\begin{document}

\title{A multiconfigurational hybrid density-functional theory}

\author{Kamal Sharkas$^{1,2}$}\email{kamal.sharkas@etu.upmc.fr}
\author{Andreas Savin$^1$}
\author{Hans J{\o}rgen Aa. Jensen$^3$}
\author{Julien Toulouse$^{1}$}\email{julien.toulouse@upmc.fr}
\affiliation{
$^1$Laboratoire de Chimie Th\'eorique, Universit\'e Pierre et Marie Curie and CNRS, 75005 Paris, France\\
$^2$Atomic Energy Commission of Syria, P.O. Box 6091, Damascus, Syria\\
$^3$Department of Physics, Chemistry, and Pharmacy, University of Southern Denmark, DK-5230 Odense M, Denmark}


\date{\today}

\begin{abstract}
We propose a multiconfigurational hybrid density-functional theory which rigorously combines a multiconfiguration self-consistent-field calculation with a density-functional approximation based on a linear decomposition of the electron-electron interaction. This gives a straightforward extension of the usual hybrid approximations by essentially adding a fraction $\l$ of exact static correlation in addition to the fraction $\l$ of exact exchange. Test calculations on the cycloaddition reactions of ozone with ethylene or acetylene and the dissociation of diatomic molecules with the Perdew-Burke-Ernzerhof (PBE) and Becke-Lee-Yang-Parr (BLYP) density functionals show that a good value of $\l$ is $0.25$, as in the usual hybrid approximations. The results suggest that the proposed multiconfigurational hybrid approximations can improve over usual density-functional calculations for situations with strong static correlation effects. 
\end{abstract}

\maketitle

\section{Introduction}

Density-functional theory (DFT)~\cite{HohKoh-PR-64} within the Kohn-Sham (KS) scheme~\cite{KohSha-PR-65} is the most widely used method for electronic-structure calculations in atomic, molecular and solid-state systems. With the usual approximate density functionals, such as generalized-gradient approximations (GGA) and hybrid approximations mixing in a fraction of Hartree-Fock (HF) exchange, DFT KS generally gives good results for situations in which the so-called dynamic electron correlation dominates the total correlation energy, but it can yield severe errors for systems with strong static correlation, i.e. with partially filled near-degenerate orbitals (see, e.g., Ref.~\onlinecite{KocHol-BOOK-01}). It has been argued that the GGA exchange density functionals actually mimic some static correlation through their self-interaction error, though in an imperfect manner (see, e.g., Refs.~\onlinecite{GriSchBae-JCP-97,Bec-JCP-03}). There is often indeed a partial cancellation of errors between the self-interaction error which tends to give too low an energy and the neglect of static correlation which gives too high an energy (see, e.g., Ref.~\onlinecite{CohMorYan-SCI-08}). Hybrid approximations have a smaller self-interaction error and are thus often worse than pure density functionals for describing systems with static correlation (see, e.g., Ref.~\onlinecite{SchZhaTru-JPCA-05}).

Several approaches have been proposed to include explicit static correlation in density-functional theory. Artificially breaking (space and spin) symmetry by unrestricted KS calculations is the simplest approach to simulate static correlation (see, e.g., Ref.~\onlinecite{Cre-MP-01}), and it often leads to reasonable potential energy surfaces but wrong spin densities. Another possible approach consists in replacing the single KS determinant by an ensemble of determinants or, equivalently, using fractional occupation numbers for the orbitals~\cite{SlaManWilWoo-PR-69,Lev-PRA-82,Lie-IJQC-83,DunMei-JCP-83,WanSch-JCP-96,SchGriBae-TCA-98,GodOrl-JCP-99,FilSha-CPL-99,FilSha-JPCA-99,FilSha-CPL-00,FilShaWoeGriPey-CPL-00,FilSha-JPCA-00,UllKoh-PRL-01,TakYamYam-IJQC-03}, but a successful and general method based on this idea is still lacking. Configuration-interaction schemes have also been proposed in which modified Hamiltonian matrix elements include information from DFT~\cite{GriWal-JCP-99,BecStaBurBla-CP-08,WuCheVoo-JCP-07,WuKadVoo-JCP-09}. A lot of approaches consist in adding to the energy of a partially correlated wave function calculation, including near-degenerate configuration state functions coupled by the full Coulombic electron-electron interaction, an energy density functional describing the missing correlation effects~\cite{LieCle-JCP-74a,LieCle-JCP-74b,ColSal-TCA-75,ColSal-TCA-79,ColSal-JCP-83,ColSal-JCP-90,Sav-IJQC-88,Sav-JCP-89,Sav-INC-91,MieStoSav-MP-97,GutSav-PRA-07,MosSan-IJQC-91,Kra-CP-92,KraCreNor-INC-92,WuSha-CPL-99,Sto-CPL-03,YinSuCheShaWu-JCTC-12,MalMcd-JPC-94,MalMcd-JPC-96,MalMcd-JPCA-97,MalMcd-CPL-98,Mcd-MP-03,BorJorNicNac-TCA-98,GraCre-CPL-00,GraCre-PCCP-00,GraCre-MP-05,TakYamYam-CPL-02,NakUkaYamTakYam-IJQC-06,YamNakUkaTakYam-IJQC-06,UkaNakYamTakYam-MP-07,GusMalLinRoo-TCA-04,PerPer-PRA-07,PerPerMorIll-JCC-07,PerPerSan-JCP-07,WeiDelGor-JCP-08,KurLawHea-MP-09}. In these last approaches, one must use a density functional which depends on the size of the multiconfigurational expansion, in order to avoid double counting of correlation between the wave function part of the calculation and the density functional. Finally, to avoiding any double counting of correlation from the beginning, it has been proposed to decompose the Coulombic electron-electron interaction into long-range and short-range components, the long-range part being treated by a method capable of describing static correlation and the short-range part being described by a density functional approximation. The methods that have been used for the long-range part are: configuration interaction~\cite{SavFla-IJQC-95,Sav-INC-96a,Sav-INC-96,LeiStoWerSav-CPL-97,PolSavLeiSto-JCP-02,SavColPol-IJQC-03,TouColSav-PRA-04}, multiconfigurational self-consistent field (MCSCF)~\cite{PedJen-JJJ-XX,FroTouJen-JCP-07,FroReaWahWahJen-JCP-09}, multireference perturbation theory~\cite{FroCimJen-PRA-10}, constrained-pairing mean-field theory~\cite{TsuScuSav-JCP-10,TsuScu-JCP-11}, and density-matrix functional theory~\cite{Per-PRA-10,RohTouPer-PRA-10,RohPer-JCP-11}.

In this work, we explore the possibility to combine MCSCF and DFT based on a simple linear decomposition of the Coulombic electron-electron interaction, in the spirit of the usual hybrid approximations, and similarly to what was recently done for constructing theoretically justified double-hybrid approximations~\cite{ShaTouSav-JCP-11}. This approach gives a straightforward multiconfigurational extension of the standard hybrid approximations, and aims at improving their description of static correlation. After deriving this multiconfigurational hybrid density-functional theory, we test this approach on situations with strong static correlation effects, namely the cycloaddition reactions of ozone with ethylene or acetylene and the dissociation of diatomic molecules, and we compare with other methods, in particular the range-separated multiconfigurational hybrid method of Ref.~\onlinecite{PedJen-JJJ-XX,FroTouJen-JCP-07,FroReaWahWahJen-JCP-09}.

\section{Theory}
\label{sec:theory}
Using the formalism of the multideterminant extension of the KS scheme (see, e.g., Refs.~\onlinecite{TouColSav-PRA-04,AngGerSavTou-PRA-05,FroTouJen-JCP-07,ShaTouSav-JCP-11}), for any coupling constant $\l$, the {\it exact} energy can be expressed as the following minimization over multideterminant wave functions $\Psi$:
\begin{eqnarray}
E &=& \min_{\Psi} \Bigl\{\bra{\Psi}\hat{T}+\hat{V}_{\text{ext}}+\l\hat{W}_{ee} \ket{\Psi}+\bar{E}_{\H xc}^{\l}[n_{\Psi}]\Bigl\}, 
\label{EminPsi}
\end{eqnarray} 
where $\hat{T}$ is the kinetic energy operator, $\hat{V}_{\text{ext}}$ is a scalar external potential operator (e.g., nuclei-electron), $\hat{W}_{ee}$ is the electron-electron interaction operator, and $\bar{E}_{\H xc}^{\l}[n_{\Psi}]$ is the {\it complement} $\l$-dependent Hartree-exchange-correlation density functional evaluated at the density coming from $\Psi$. The complement density functional, $\bar{E}_{\H xc}^{\l}[n]=E_{\H xc}[n]-E_{\H xc}^{\l}[n]$, is the difference between the usual KS density functional $E_{\H xc}[n]$ and the $\l$-dependent density functional $E_{\H xc}^{\l}[n]$ corresponding to the interaction $\l\hat{W}_{ee}$. The Hartree-exchange contribution is of first order in the electron-electron interaction and is thus linear in $\l$,
\begin{eqnarray}
\bar{E}_{\H x }^{\l}[n]=(1-\l) E_{\H x}[n],
\label{}
\end{eqnarray} 
where $E_{\H x}[n]$ is the usual KS Hartree-exchange density functional. The correlation contribution is obtained by uniform coordinate scaling of the density~\cite{LevPer-PRA-85,LevYanPar-JCP-85,Lev-PRA-91,LevPer-PRB-93},
\begin{eqnarray}
\bar{E}_{c}^{\l}[n] &=& E_{c}[n]-E_{c}^{\l}[n]
\nonumber\\
                    &=& E_{c}[n]-\l^2 E_{c}[n_{1/\l}],
\label{Ecln}
\end{eqnarray} 
where $E_{c}[n]$ is the usual KS correlation functional, $E_{c}^{\l}[n]$ is the correlation functional corresponding to the interaction $\l\hat{W}_{ee}$, and $n_{1/\l}(\b{r})=(1/\l)^3 n(\b{r}/\l)$ is the scaled density.

The theory is so far exact but in practice approximations must be used for the multideterminant wave function and the density functionals. In Ref.~\onlinecite{ShaTouSav-JCP-11}, by restricting the search in Eq.~(\ref{EminPsi}) to single-determinant wave functions $\Phi$, we defined the \textit{density-scaled one-parameter hybrid} (DS1H) approximation
\begin{eqnarray}
E^{\text{DS1H}} = \min_{\Phi} \Bigl\{ \bra{\Phi}\hat{T}+\hat{V}_{\text{ext}}+\l\hat{W}_{ee} \ket{\Phi} \;\;\;\;\;\;\;\;\;\;\;\;\;\;\;
\nonumber\\
+ (1-\l) E_{\H x}[n_\Phi] + E_{c}[n_\Phi]-\l^2 E_{c}[n_{\Phi,1/\l}] \Bigl\}, 
\end{eqnarray} 
and, by additionally neglecting density scaling in the correlation functional, $E_{c}[n_{1/\l}] \approx E_{c}[n]$, we obtained the \textit{one-parameter hybrid} (1H) approximation,
\begin{eqnarray}
E^{\text{1H}} = \min_{\Phi} \Bigl\{ \bra{\Phi}\hat{T}+\hat{V}_{\text{ext}}+\l\hat{W}_{ee} \ket{\Phi} \;\;\;\;\;\;\;\;\;\;\;\;\;\;\;
\nonumber\\
+ (1-\l) E_{\H x}[n_\Phi] + (1 -\l^2) E_{c}[n_\Phi] \Bigl\}, 
\end{eqnarray} 
which is similar to the usual one-parameter hybrid approximations~\cite{Bec-JCP-96,ErnPerBur-INC-96}, except that the correlation functional is weighted by a factor of $(1 -\l^2)$. Starting from these references and applying a second-order M{\o}ller-Plesset (MP2) perturbation theory~\cite{AngGerSavTou-PRA-05,FroJen-PRA-08,Ang-PRA-08}, we also defined the \textit{density-scaled one-parameter double-hybrid} (DS1DH) and \textit{one-parameter double-hybrid} (1DH) approximations~\cite{ShaTouSav-JCP-11}, which are one-parameter versions of the original double-hybrid approximations~\cite{Gri-JCP-06}. These latter also combine HF exchange and MP2 correlation with a semilocal exchange-correlation density functional but with two empirical parameters.

\begin{figure*}
\includegraphics[scale=0.3,angle=-90]{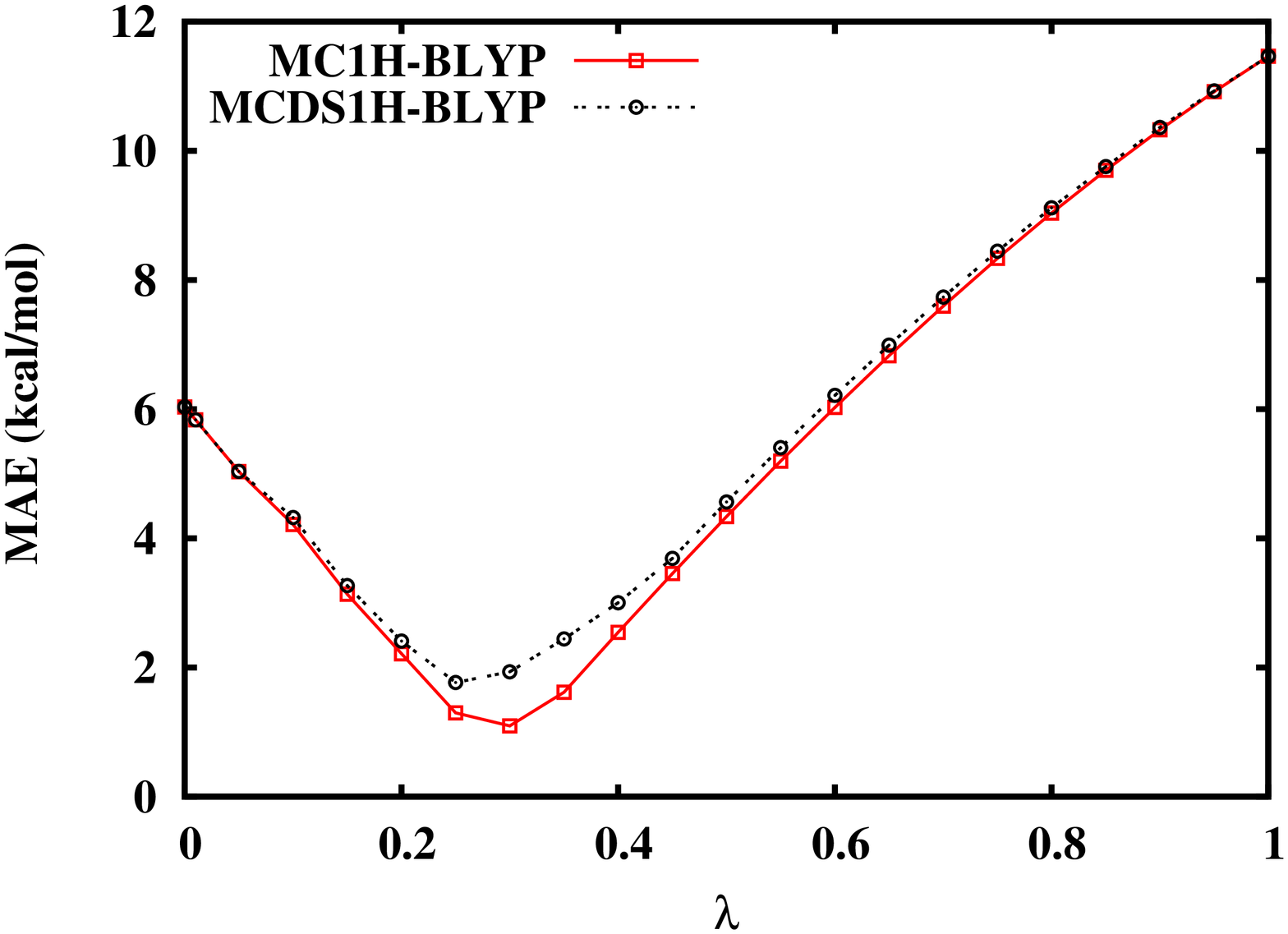}
\includegraphics[scale=0.3,angle=-90]{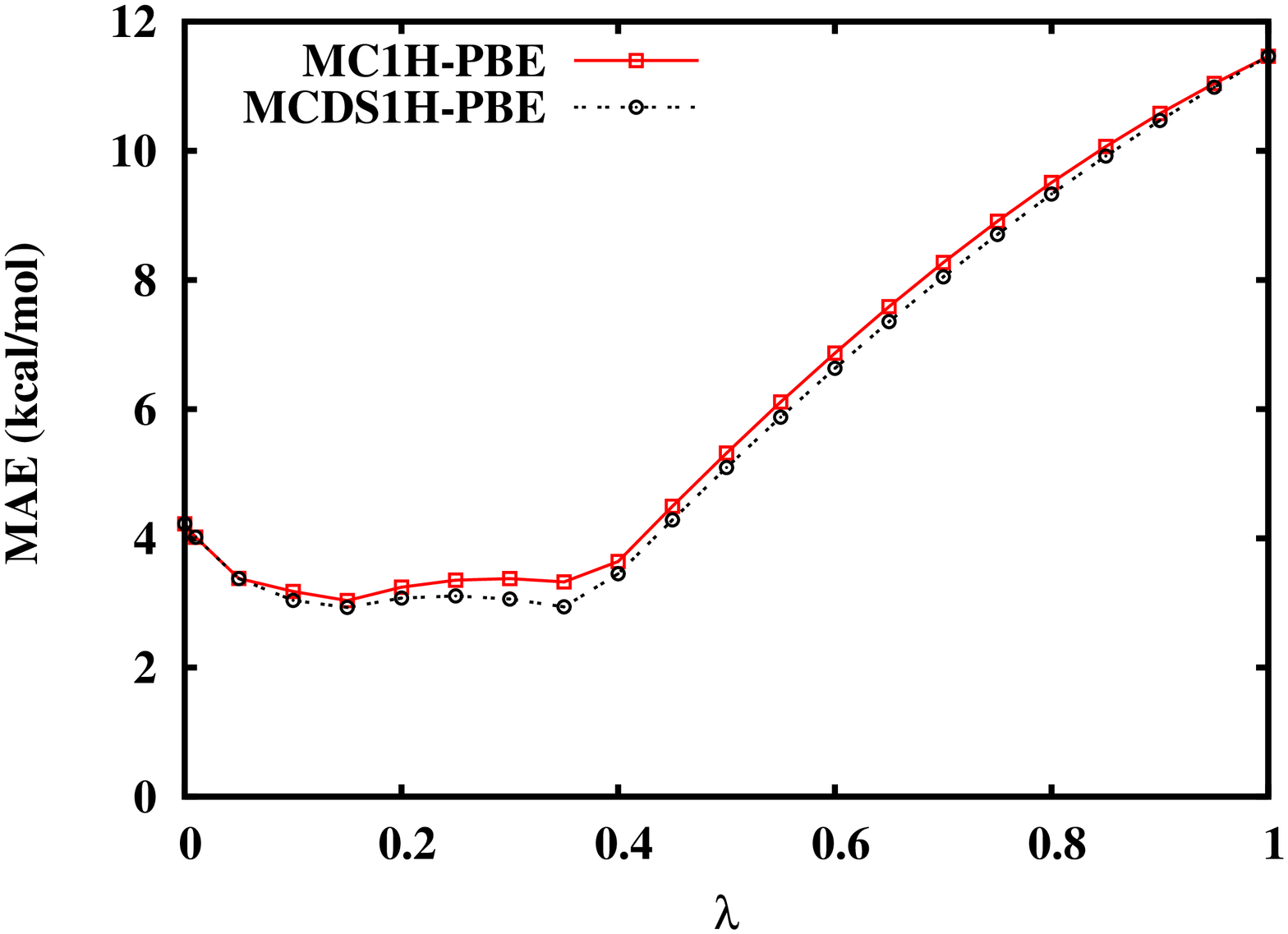}
\caption{MAEs of O3ADD6 set as functions of the parameter $\l$ for the MC1H and MCDS1H approximations with the BLYP (left) and PBE (right) exchange-correlation density functionals. All calculations were carried out with the aug-cc-pVTZ basis set.}
\label{fig:BHO3ADD6}
\end{figure*}

Here, we follow a different route and use an MCSCF wave function in Eq.~(\ref{EminPsi}), expanded as a linear combination of configuration state functions $\Phi_I$,
\begin{equation}
\ket{\Psi} = \sum_I c_I \ket{\Phi_I},
\label{Psi}
\end{equation}
where the coefficients $c_I$ and the orbitals are to be simultaneously optimized. With this form of wave function, we obtain a \textit{multiconfigurational density-scaled one-parameter hybrid} (MCDS1H) approximation,
\begin{eqnarray}
E^{\text{MCDS1H}} = \min_{\Psi} \Bigl\{ \bra{\Psi}\hat{T}+\hat{V}_{\text{ext}}+\l\hat{W}_{ee} \ket{\Psi} \;\;\;\;\;\;\;\;\;\;\;\;\;\;\;
\nonumber\\
+ (1-\l) E_{\H x}[n_\Psi] + E_{c}[n_\Psi]-\l^2 E_{c}[n_{\Psi,1/\l}] \Bigl\}, 
\label{MCDS1H}
\end{eqnarray} 
and, if density scaling, which is not considered in usual hybrid approximations, is neglected, we obtain a \textit{multiconfigurational one-parameter hybrid} (MC1H) approximation,
\begin{eqnarray}
E^{\text{MC1H}} = \min_{\Psi} \Bigl\{ \bra{\Psi}\hat{T}+\hat{V}_{\text{ext}}+\l\hat{W}_{ee} \ket{\Psi} \;\;\;\;\;\;\;\;\;\;\;\;\;\;\;
\nonumber\\
+ (1-\l) E_{\H x}[n_\Psi] + (1 -\l^2) E_{c}[n_\Psi] \Bigl\}. 
\label{MC1H}
\end{eqnarray} 
Equation~(\ref{MCDS1H}) is the equivalent of the range-separated multiconfigurational hybrid method of Refs.~\onlinecite{PedJen-JJJ-XX,FroTouJen-JCP-07,FroReaWahWahJen-JCP-09}, that we will refer to as MC-srDFT, but for a linear separation of the electron-electron interaction. Notice that, if we were to use no approximations for the wave function $\Psi$ and the exchange-correlation density functional, then Eq.~(\ref{MCDS1H}) would give the exact energy, {\it independently} of $\lambda$. In practice, of course, we must use approximations, and the energy does depend on $\l$, which can then be considered as an empirical parameter to be optimized.

The present scheme has two advantages over the range-separated scheme: (a) Only one list of two-electron Coulomb integrals is needed and it is just multiplied by $\l$ in the MCSCF part and by $(1-\l)$ in the complement Hartree energy, whereas two lists of two-electron integrals are needed in the range-separated scheme for the long-range MCSCF part and for the short-range complement Hartree energy; (b) No new exchange and correlation density functionals need in principle to be developed since all the existing approximations developed for the KS scheme can be reused with a simple scaling, whereas new short-range density-functional approximations must be developed in the range-separated scheme.

Equations~(\ref{MCDS1H}) and~(\ref{MC1H}) can be seen as straightforward multiconfigurational extensions of the usual hybrid approximations. Indeed, the expectation value of $\lambda \hat{W}_{ee}$ over the MCSCF wave function $\Psi$ not only introduces a fraction $\l$ of exact exchange but also a fraction of exact static correlation. Defined in the ideal limit of reference energy levels that are degenerate, the static correlation energy is linear with respect to the electron-electron interaction, and thus we can consider that the expectation value of $\lambda \hat{W}_{ee}$ introduces a linear fraction $\l$ of static correlation. By contrast, the dynamic correlation energy starts at quadratic order in the electron-electron interaction, so that, for sufficiently small $\l$, it is justified to neglect it in the wave function expectation value. Moreover, for sufficiently small $\l$, the weight $(1-\l^2)$ is close to $1$ and thus Eq.~(\ref{MC1H}) includes a nearly complete approximate correlation energy functional, that is often thought of as correctly describing dynamic correlation. Of course, if the multiconfigurational hybrid approximations of Eq.~(\ref{MCDS1H}) or~(\ref{MC1H}) are to be accurate, the fraction $(1-\l)$ of static correlation energy not treated by MCSCF must be accounted for by the density functional, possibly through a compensation with the self-interaction error of the scaled exchange functional $(1-\l) E_{x}[n]$.

\begingroup
\begin{table*}[t]
\caption{Energies of the van der Waals (vdW) complex, the transition state (TS), and the cycloadduct (primary ozonide), relative to the separated reactants, and the corresponding MAEs (in kcal/mol) for the addition of ozone with acetylene or ethylene (O3ADD6 set), calculated by several methods. For the DS1DH-PBE and 1DH-BLYP double-hybrid approximations, we use the value $\l=0.65$ which was previously optimized in Ref.~\onlinecite{ShaTouSav-JCP-11}. For the MCDS1H and MC1H multiconfigurational hybrid approximations, we use a value of $\l=0.25$ which roughly minimizes the MAE according to Fig.~\ref{fig:BHO3ADD6}. For the range-separated MC-srPBE multiconfigurational hybrid approximation, we use the value of the range-separation parameter $\mu=0.40$ bohr$^{-1}$ which was previously determined in Ref.~\onlinecite{FroTouJen-JCP-07}. For the multiconfigurational methods, a complete active space wave function with 2 electrons in 2 orbitals [CAS(4,4)] is chosen for the van der Waals complex, the transition state, and the cycloadduct, and a CAS(2,2) wave function for each separate reactant. All calculations were carried out with the aug-cc-pVTZ basis set. All calculations were done for M05/6-311+G(2df,2p) geometries, except for the MRMP2 results which are for CCSD(T)/cc-pVTZ geometries.}
\label{tab:O3ADD6}
\begin{tabular}{lcccccccccr}
\hline
\hline
   &    & \multicolumn{3}{c}{O$_{3}$ + C$_{2}$H$_{2}$ $ \xrightarrow ~ $}    & & \multicolumn{3}{c}{O$_{3}$ + C$_{2}$H$_{4}$ $ \xrightarrow ~ $}  \\
             \cline{3-5} \cline{7-9}
Method      &   &  vdW    &    TS &        cycloadduct &   \phantom{xxxx} & vdW     &    TS  &       cycloadduct      &&   \phantom{xxxx} MAE   \\
\hline
HF          &    &0.68       &23.08     & -87.12 &                 & 1.90      & 17.91     & -82.58      && 14.18\\
MCSCF       &    &0.69       &27.54     & -77.25 &                 & 1.32      &22.13      &-68.06       && 11.46      \\
MP2         &    & -3.18     & 1.13     & -54.81 &                 & -4.01     &-5.74      &-51.18       && 5.67      \\
MRMP2$^a$   &    &-2.16      & 8.77     &-48.19  &                 &-2.09      &3.43       &-43.32       && 5.16        \\
PBE         &    &-1.71      &-1.66     & -62.44 &                 &-2.50      & -4.77     &-51.45       && 4.23    \\
BLYP        &    &-0.57      &2.21      &-53.98  &                 &-1.29      & -1.55     &-43.19       && 6.03     \\
\\
Single-hybrid approximations\\
PBE0        &    &-1.26      &1.65      &-74.00  &                 &-1.55      & -1.74     &-64.74       && 5.00     \\
B1LYP       &    &-0.60      &4.78      &-66.56  &                 &-0.83      &0.71       &-57.47       && 1.85     \\
B3LYP       &    &-0.67      &3.81      &-65.10  &                 &-1.01      &-0.12      &-55.64       && 2.06     \\
\\
Double-hybrid approximations\\
DS1DH-PBE   &    & -2.08     & 3.47     & -61.21 &                 & -2.54     &-1.81      &-54.90       && 2.51      \\
1DH-BLYP    &    & -1.92     & 4.42     & -58.37 &                 & -2.37     &-1.14      &-52.12       && 3.12      \\
B2-PLYP$^b$  &    &-1.47      & 5.00     &-60.18  &                 & -1.81     &-0.13      &-53.05       && 2.42      \\
\\
Multiconfigurational hybrid approximations\\
MC-srPBE    &    &-0.93      &4.12      &-72.73  &                 &-0.87      & 0.71      &-65.53       && 4.27     \\
MCDS1H-PBE  &    &-1.06      &3.88      &-70.41  &                 &-1.22      & 0.30      &-60.71       && 3.11     \\
MC1H-PBE    &    &-1.08      &3.66      &-70.97  &                 &-1.25      & 0.13      &-61.26       && 3.35     \\
MCDS1H-BLYP &    & 0.28      &7.94      &-62.47  &                 &0.26       & 3.78      &-52.86       && 1.77     \\
MC1H-BLYP   &    &-0.36      &6.74      &-63.76  &                 &-0.47      &2.57       &-54.21       && 1.30    \\
\\
Best estimate$^a$ & &-1.90      &7.74         &-63.80 &                  &-1.94      &3.37      &-57.15         &&       \\
\hline
\hline
$^a$From Ref.~\onlinecite{ZhaTisGouLiLutPieTru-JPCA-09}.\\
$^b$Performed with the Gaussian09 program~\cite{Gaussian-PROG-09}.
\end{tabular}
\end{table*}
\endgroup

\section{Computational details}

The calculations have been performed with a development version of the DALTON 2011 program~\cite{Dal-PROG-11}, in which the MCDS1H and MC1H approximations have been implemented in the same way than for the MC-srDFT method~\cite{Ped-THESIS-04,PedJen-JJJ-XX,FroTouJen-JCP-07}, using the direct restricted-step second-order MCSCF algorithm of Jensen and coworkers~\cite{JenJor-JCP-84,JenAgr-CPL-84,JenAgr-CP-86,JenJorAgr-JCP-87,JenJorAgrOls-JCP-88,Jen-INC-94}. For $E_{x}[n]$ and $E_{c}[n]$, we use two GGA exchange-correlation density functionals, Perdew-Burke-Ernzerhof (PBE)~\cite{PerBurErn-PRL-96} and Becke/Lee-Yang-Parr (BLYP)~\cite{Bec-PRA-88,LeeYanPar-PRB-88}, without spin-density dependence. For implementing the density-scaled correlation functionals in the MCSCF algorithm, we need the scaling relations for the energy density, and its first- and second-order derivatives that we give in Appendix~\ref{app:scaling}. The computational cost of the method is essentially the same as for a standard MCSCF calculation, with a small extra cost due to the DFT contribution.

A good value for the empirical parameter $\l$ in Eqs.~(\ref{MCDS1H}) and (\ref{MC1H}) is determined on the O3ADD6 benchmark set~\cite{GoeGri-JCTC-10,GoeGri-JCTC-11} for the 1,3-dipolar cycloaddition reactions of ozone (O$_3$) with ethylene (C$_2$H$_4$) or acetylene (C$_2$H$_2$)~\cite{WheEssHou-JPCA-07,ZhaTisGouLiLutPieTru-JPCA-09,SaiNisKatNakKitKawYamOkuYam-JPCA-10}. For these two reactions, there are three stationary points along the reaction coordinate: the van der Waals complex, the transition state, and the cycloadduct (primary ozonide), all in a closed-shell spin-singlet state. The O3ADD6 set consists of the six energies of these stationary points of the two reactions, calculated relative to the energy of the separated reactants, and without zero-point vibrational energy correction. Accurate calculations of these energies are difficult and require to handle the subtle balance between static and dynamic correlation effects along the reaction coordinate. The ozone reactant, the van der Waals complex, and the transition state have a strong multiconfigurational character corresponding to the HOMO $\to$ LUMO double excitation in ozone. In the cycloadduct, and to a less extent in the transition state, the stabilization of the ozone HOMO and the destabilization of the ozone LUMO greatly reduce this multiconfigurational character. In addition, there are small near-degeneracy correlation effects due to the $\pi$ and $\pi^{*}$ orbitals of the reactive $\pi$ bond of ethylene and acetylene. For each separate reactant, a CAS(2,2) wave function is chosen, the active space corresponding to the HOMO and LUMO orbitals for ozone, and to the HOMO ($\pi$) and LUMO ($\pi^{*}$) orbitals of the reactive $\pi$ bond for ethylene and acetylene. For the van der Waals complex, the transition state, and the cycloadduct, a CAS(4,4) wave function is consistently chosen, the active space corresponding to the orbitals that connect to the ones chosen for the reactants in the dissociation limit. We use the aug-cc-pVTZ basis set~\cite{Dun-JCP-89,KenDunHar-JCP-92} and the fixed geometries of Ref.~\onlinecite{ZhaTisGouLiLutPieTru-JPCA-09} optimized using the hybrid meta-GGA exchange-correlation functional M05~\cite{ZhaSchTru-JCP-05} with the 6-311+G(2df,2p) basis set~\cite{KriBinSeePop-JCP-80,MclCha-JCP-80}. The reference values for the energies are from Ref.~\onlinecite{ZhaTisGouLiLutPieTru-JPCA-09} and were obtained from extensive coupled-cluster calculations extrapolated to the complete basis set limit~\cite{WheEssHou-JPCA-07,ZhaTisGouLiLutPieTru-JPCA-09}. We calculate the mean absolute error (MAE) over the six values as a function of the parameter $\l$. We compare the MCDS1H and MC1H approximations with (a) some non-hybrid methods:  HF, MCSCF, MP2~\cite{MolPle-PR-34}, multireference MP2 (MRMP2)~\cite{Hir-CPL-92,ZhaTisGouLiLutPieTru-JPCA-09}, PBE~\cite{PerBurErn-PRL-96}, and BLYP~\cite{Bec-PRA-88,LeeYanPar-PRB-88}; (b) some single-hybrid approximations: PBE0~\cite{AdaBar-JCP-99,AdaBar-CPL-98}, B1LYP~\cite{AdaBar-CPL-97}, and B3LYP~\cite{Bec-JCP-93,BarAda-CPL-94}; (c) some double-hybrid approximations: DS1DH-PBE~\cite{ShaTouSav-JCP-11}, 1DH-BLYP~\cite{ShaTouSav-JCP-11}, and B2-PLYP~\cite{Gri-JCP-06}, all applied in a spin-restricted formalism. We also compare with the range-separated MC-srPBE multiconfigurational hybrid approximation~\cite{PedJen-JJJ-XX,FroTouJen-JCP-07,FroReaWahWahJen-JCP-09} using the short-range PBE exchange-correlation functional of Ref.~\onlinecite{GolWerStoLeiGorSav-CP-06} and the value of the range-separation parameter $\mu=0.40$ bohr$^{-1}$ which was previously determined in Ref.~\onlinecite{FroTouJen-JCP-07}.

We also test the MCDS1H and MC1H approximations by computing the potential energy curves the five diatomic molecules H$_{2}$, Li$_{2}$, C$_{2}$, N$_{2}$, and F$_{2}$, using in each case a full-valence CAS wave function and the cc-pVTZ basis set~\cite{Dun-JCP-89}.

\begin{figure*}
\includegraphics[scale=0.3,angle=-90]{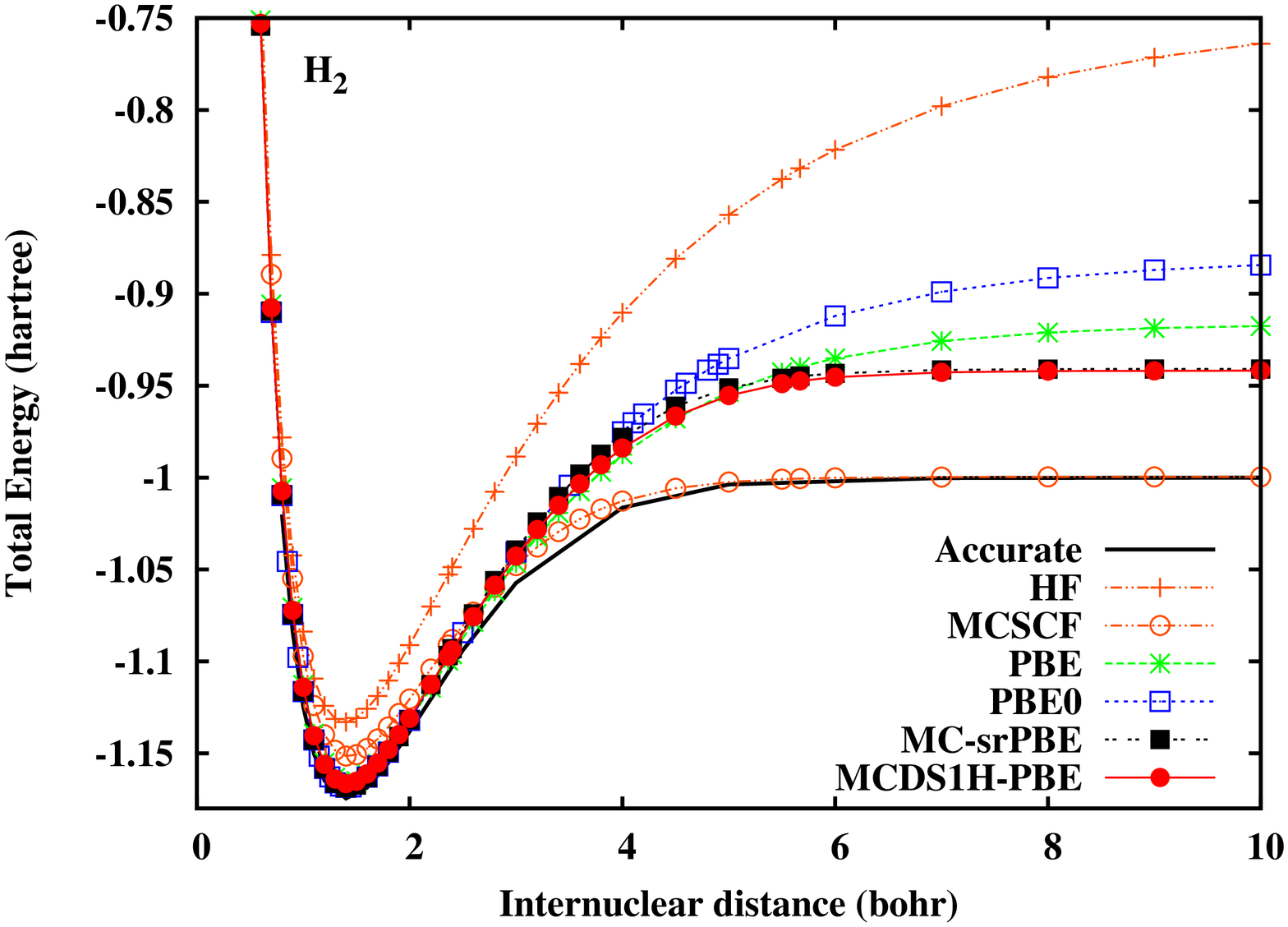}
\includegraphics[scale=0.3,angle=-90]{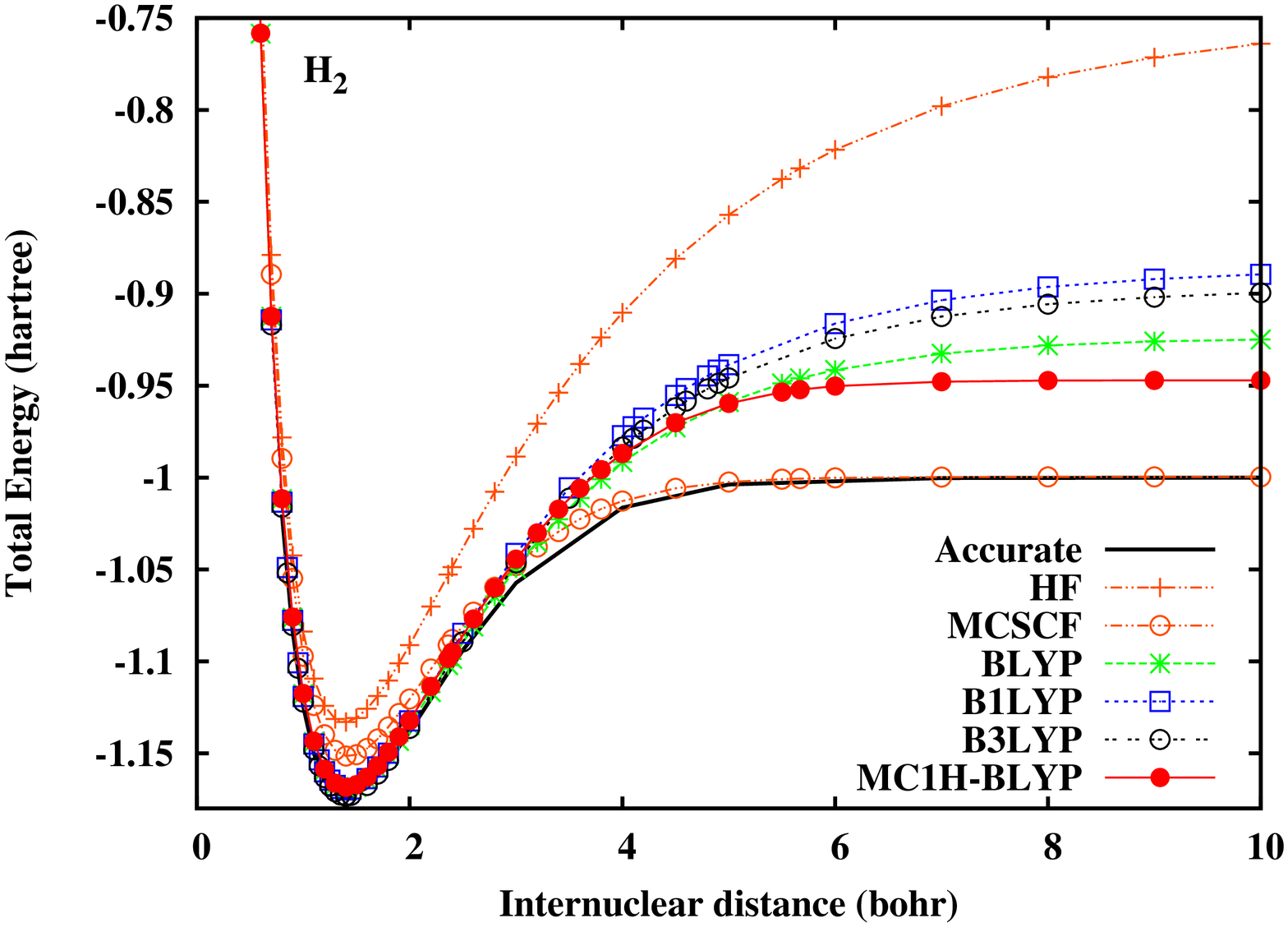}
\caption{Potential energy curves of H$_2$ calculated by HF, MCSCF, and several methods based on the PBE (left) or BLYP (right) exchange-correlation density functionals. For the MCDS1H and MC1H multiconfigurational hybrid approximations, we use a value of $\l=0.25$. For the range-separated MC-srPBE multiconfigurational hybrid approximation, we use a value of the range-separation parameter of $\mu=0.40$ bohr$^{-1}$. For all multiconfigurational methods, we use a full-valence CAS wave function. The basis set is cc-pVTZ basis. The accurate curve is from Ref.~\onlinecite{LieCle-JCP-74b}.}
\label{fig:h2}
\end{figure*}

\section{Results}

\subsection{O3ADD6 database}
\label{sec:O3ADD6}

Figure~\ref{fig:BHO3ADD6} shows the MAEs for the O3ADD6 set as functions of the parameter $\l$ for the MCDS1H and MC1H approximations with the BLYP and PBE exchange-correlation density functionals. For $\l=0$, the MCDS1H and MC1H approximation reduce to a standard KS calculation with the corresponding approximate density functional. For $\l=1$, they reduce to a standard MCSCF calculation. Contrary to what was found for the calculation of atomization energies using double-hybrid approximations~\cite{ShaTouSav-JCP-11}, here neglecting density scaling in the correlation functional makes little difference. Toward the $\l=1$ end of the curves, the MCDS1H and MC1H approximations inherit the inaccuracy of MCSCF which neglects dynamic correlation. The MAE curves of the MCDS1H-BLYP and MC1H-BLYP approximations display a marked minimum at an intermediate value of $\l$, thus improving upon both the standard BLYP and MCSCF calculations. The minimum is reached at $\l=0.25$ for MCDS1H-BLYP and at $\l=0.30$ for MC1H-BLYP, with MAEs below 2 kcal/mol. For the MCDS1H-PBE and MC1H-PBE approximations, the MAE curves have a plateau around $\l=0.25$ with a MAE of about 3 kcal/mol, which is again smaller than both the standard PBE and MCSCF calculations. In view of these results, we choose the value $\l=0.25$ in all MCDS1H and MC1H approximations. It is a conservative choice since it gives the same fraction of exact exchange as the one usually advocated in the usual single-hybrid approximations~\cite{PerErnBur-JCP-96}.

Table~\ref{tab:O3ADD6} reports the energies of the van der Waals complex, the transition state, and the cycloadduct of the two reactions of the O3ADD6 set, relative to the separated reactants, calculated with the MCDS1H and MC1H approximations at $\l=0.25$. For comparison, we also report results for various non-hybrid and other hybrid methods.

During the early stages of the two reactions, a weakly bound van der Waals complex is formed which lies in a shallow minimum (-1.90 kcal/mol for C$_2$H$_2$ and -1.94 kcal/mol for C$_2$H$_4$) below the reactants. The MCDS1H and MC1H approximations give significantly underestimated well depths, which are still in improvement over standard MCSCF but not over standard KS calculations with the corresponding functionals. The range-separated MC-srPBE method does also not perform better than KS PBE for these van der Waals systems. A better description of the long-range dispersion correlation would indeed require inclusion of perturbation corrections on top of the active space~\cite{FroCimJen-PRA-10}. As expected, the double-hybrid approximations, which include second-order perturbation corrections, tend to perform better for these van der Waals complexes. The best performance is achieved with MRMP2 which is able to correctly describe both multiconfigurational effects and dispersion correlations. However, one should keep in mind than the methods MP2, MRMP2, and the double hybrids are most likely less converged with respect to the basis size than the other methods.

The activation barriers of the transition states are underestimated (or not present at all) in KS PBE and BLYP calculations, and to a less extent with the single-hybrid and double-hybrid approximations, while they are largely overestimated in MCSCF. Note that here, contrary to the common case, MCSCF gives higher activation barriers than HF because the ozone reactant has more static correlation than the transition state. The MCDS1H and MC1H approximations give an improvement of about 5 kcal/mol over the KS calculations with the corresponding functionals. The range-separated MC-srPBE method gives activation barriers which are slightly better than the ones given by MCDS1H-PBE and MC1H-PBE, but largely worse than the ones given by MCDS1H-BLYP and MC1H-BLYP. The values obtained with MCDS1H-BLYP and MC1H-BLYP, as well as with MRMP2, are all within 1 kcal/mol of the best estimates.

\begin{figure*}
\includegraphics[scale=0.3,angle=-90]{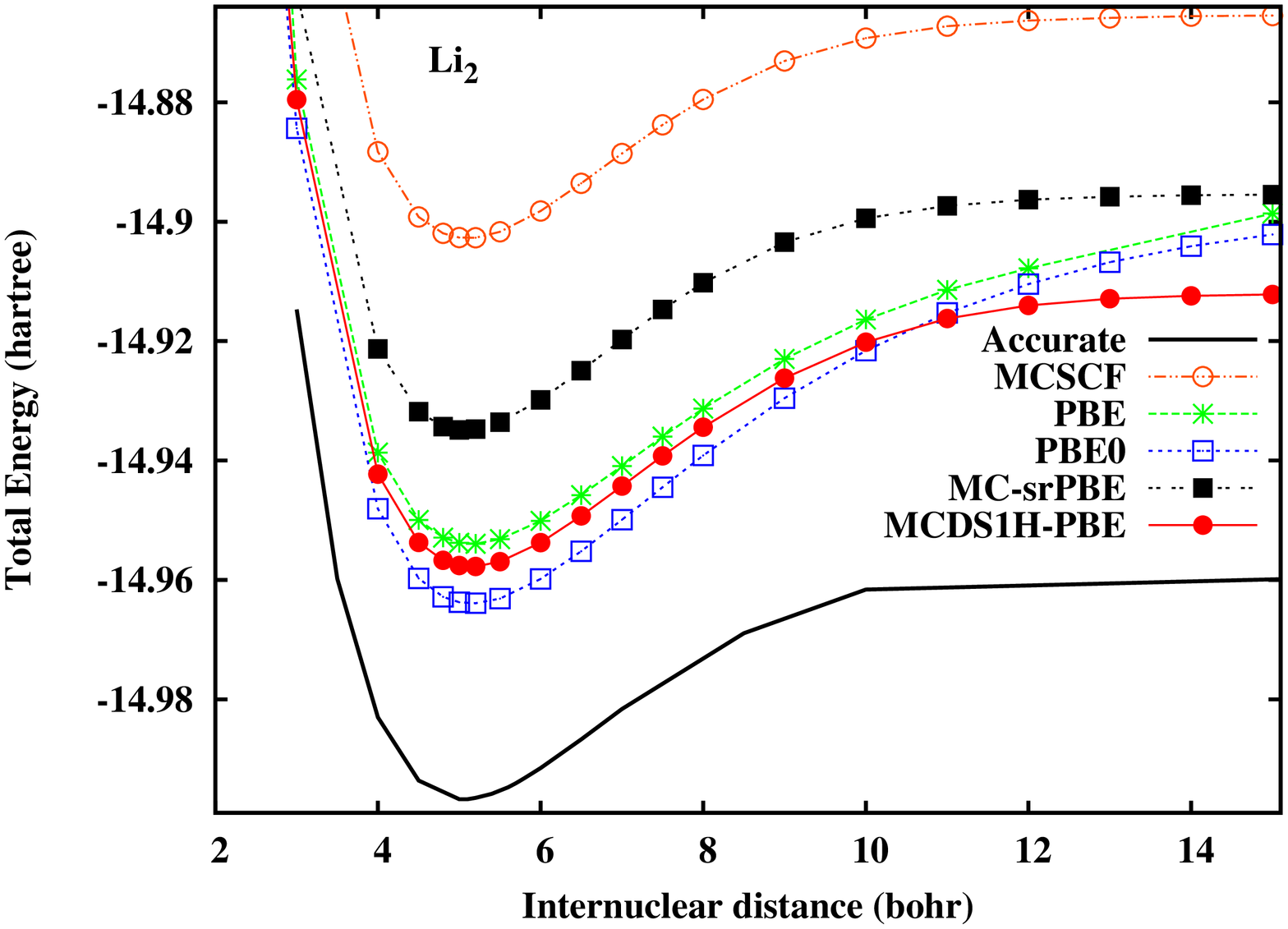}
\includegraphics[scale=0.3,angle=-90]{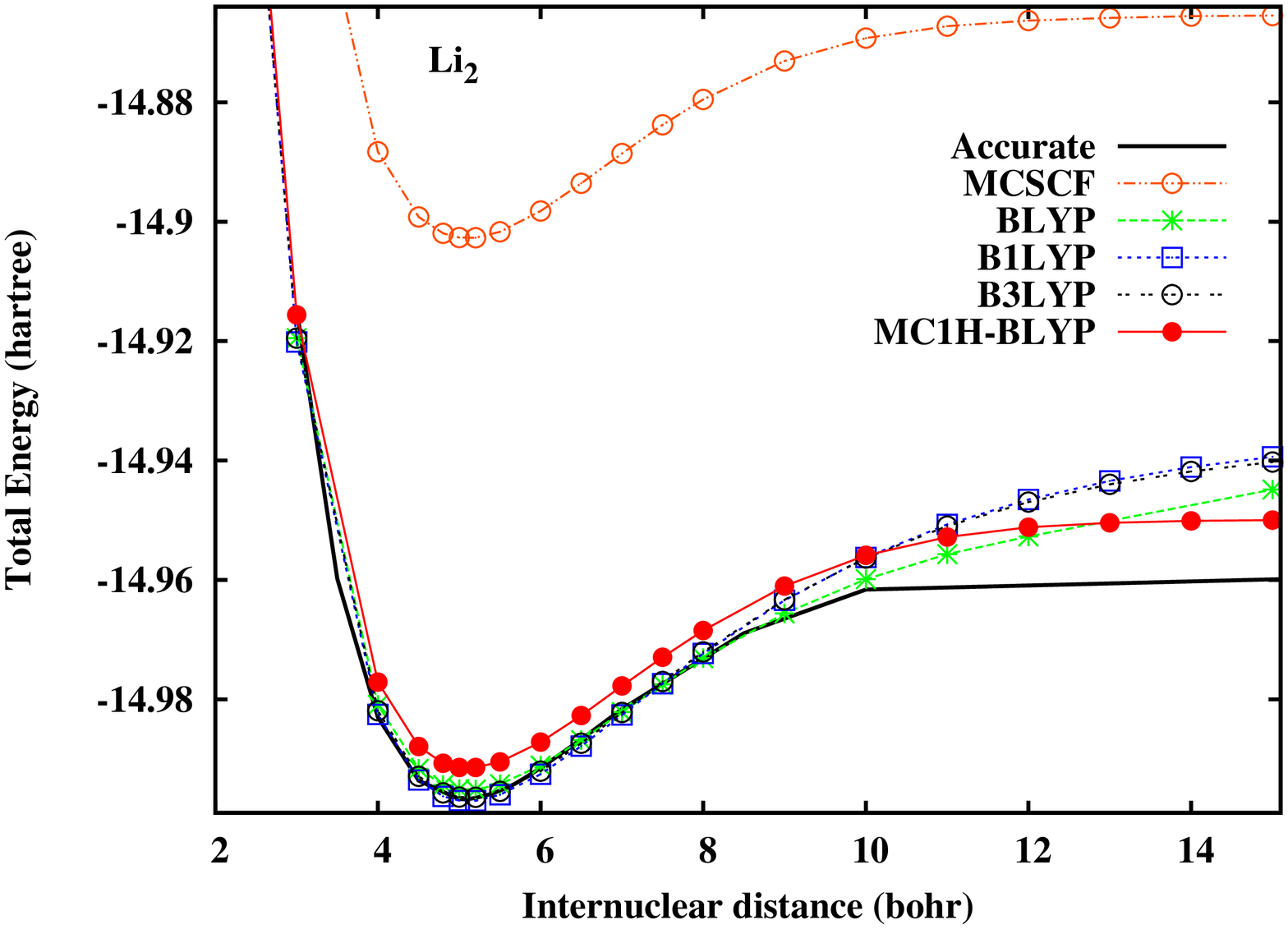}
\includegraphics[scale=0.3,angle=-90]{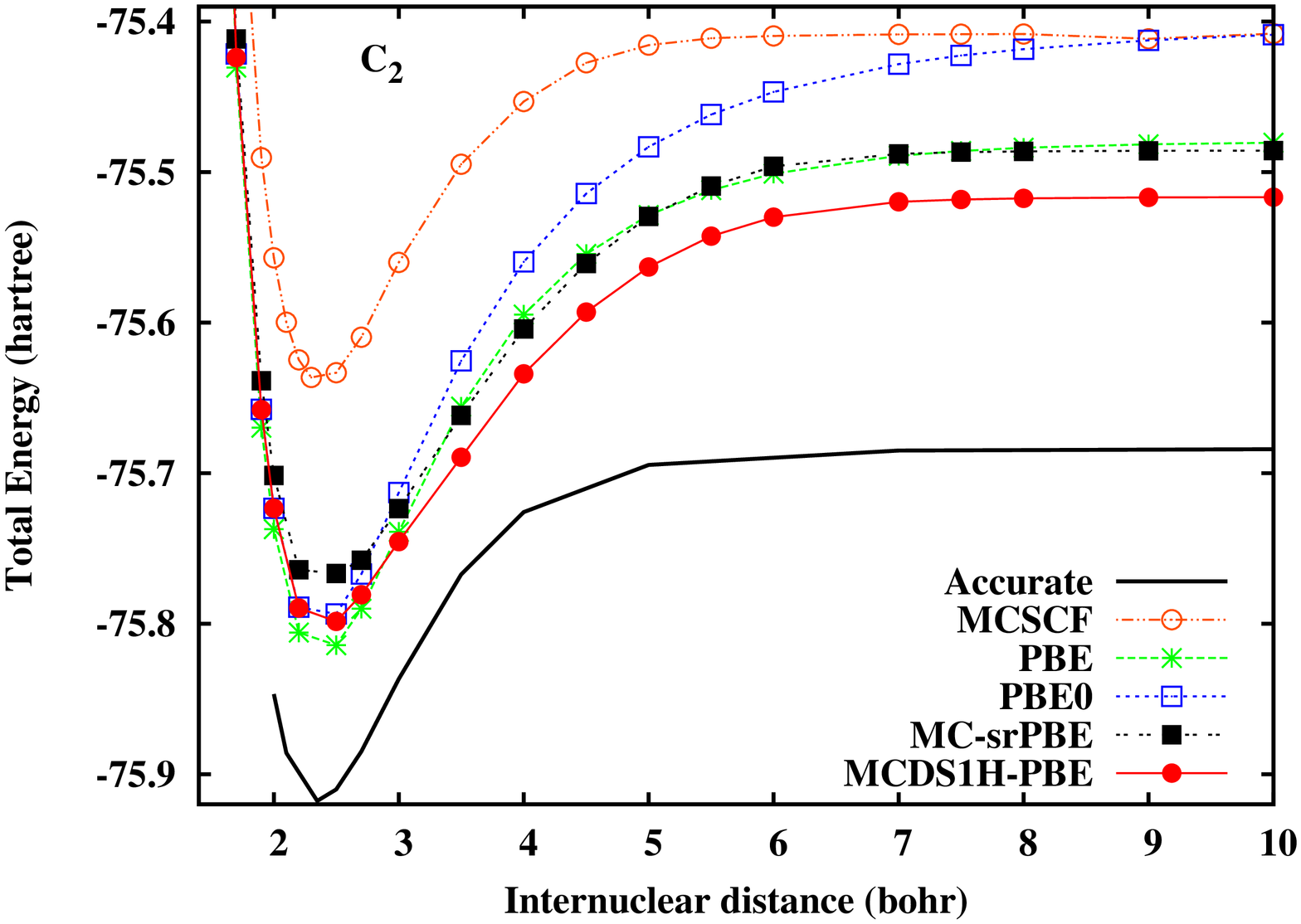}
\includegraphics[scale=0.3,angle=-90]{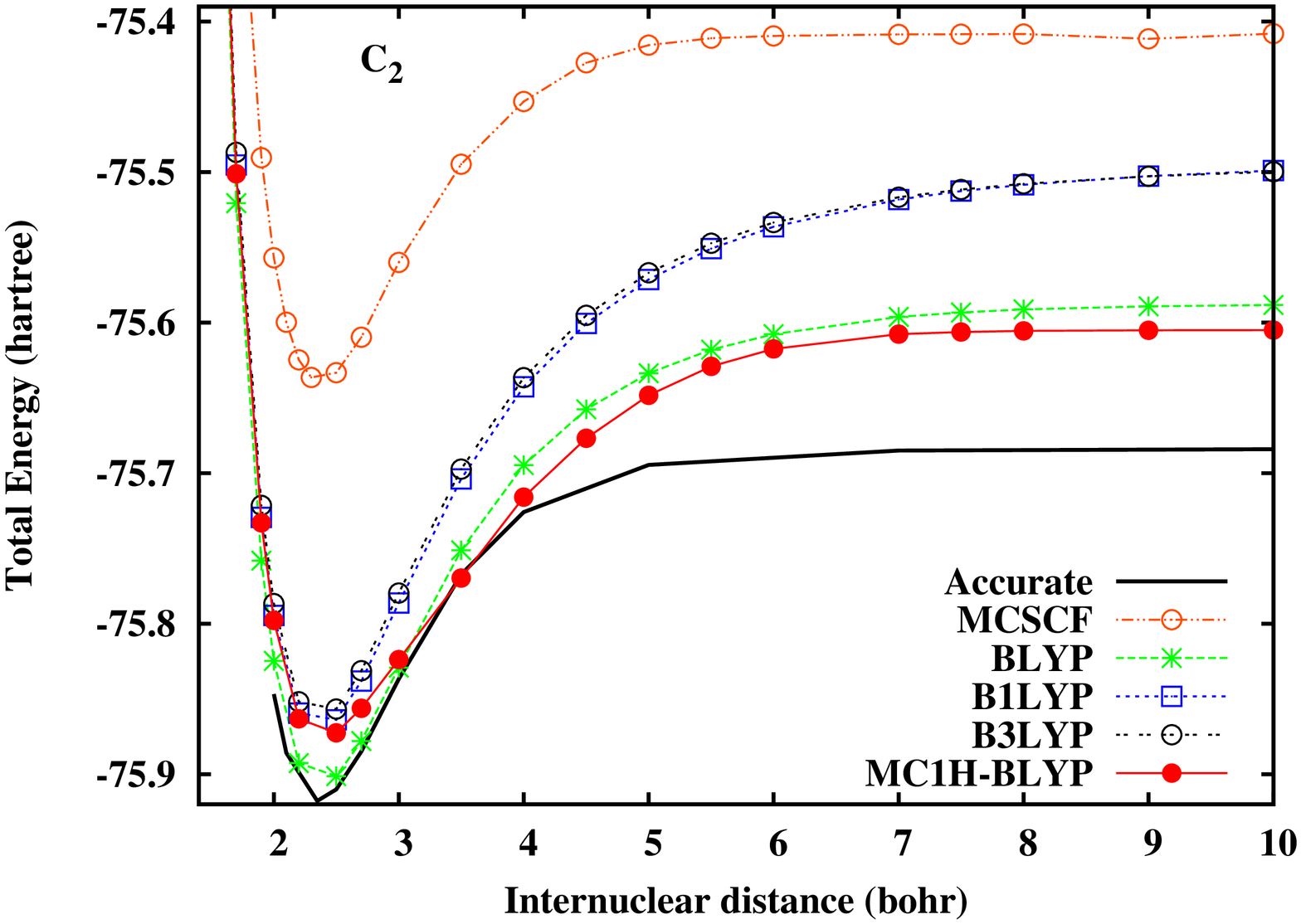}
\includegraphics[scale=0.3,angle=-90]{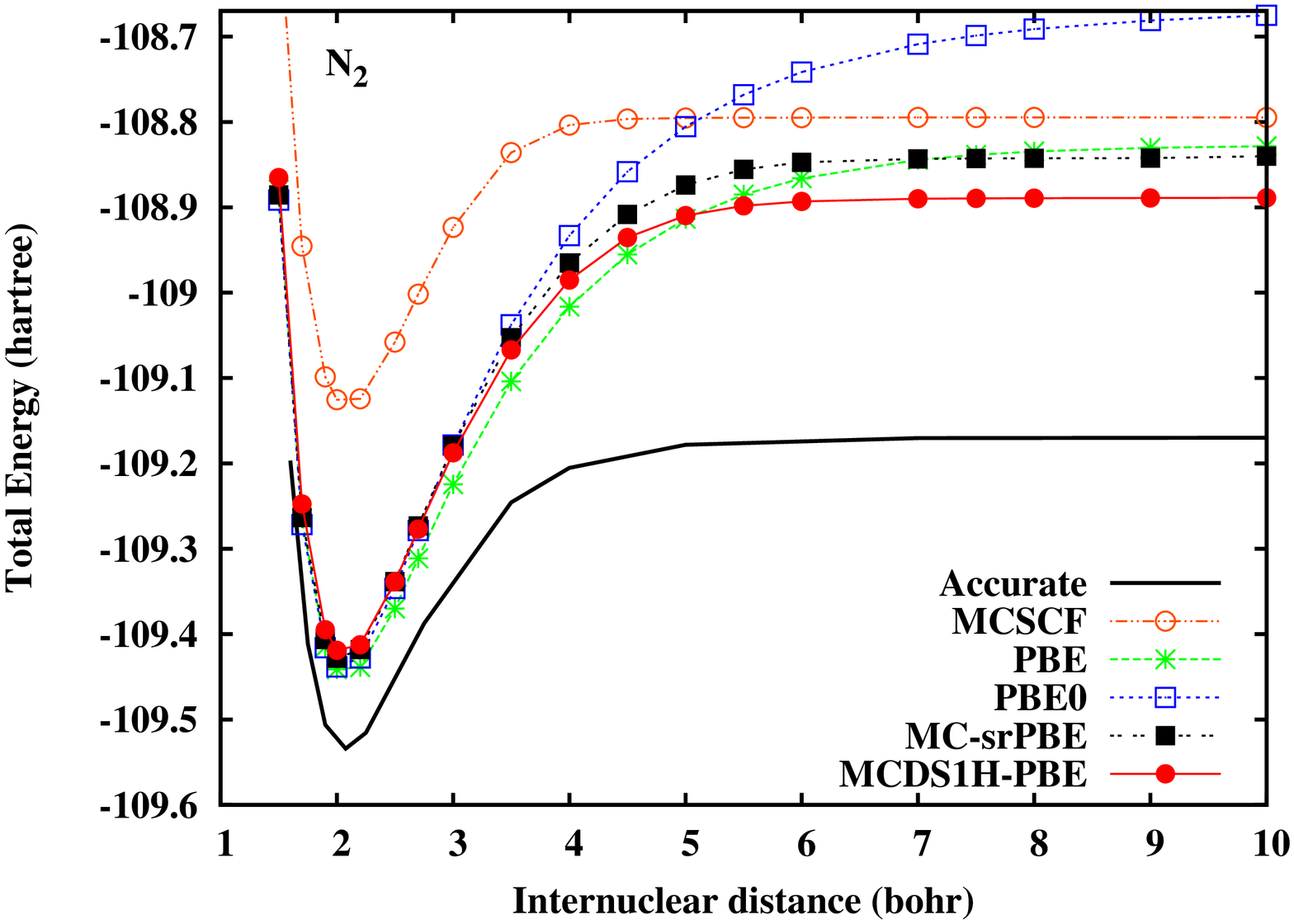}
\includegraphics[scale=0.3,angle=-90]{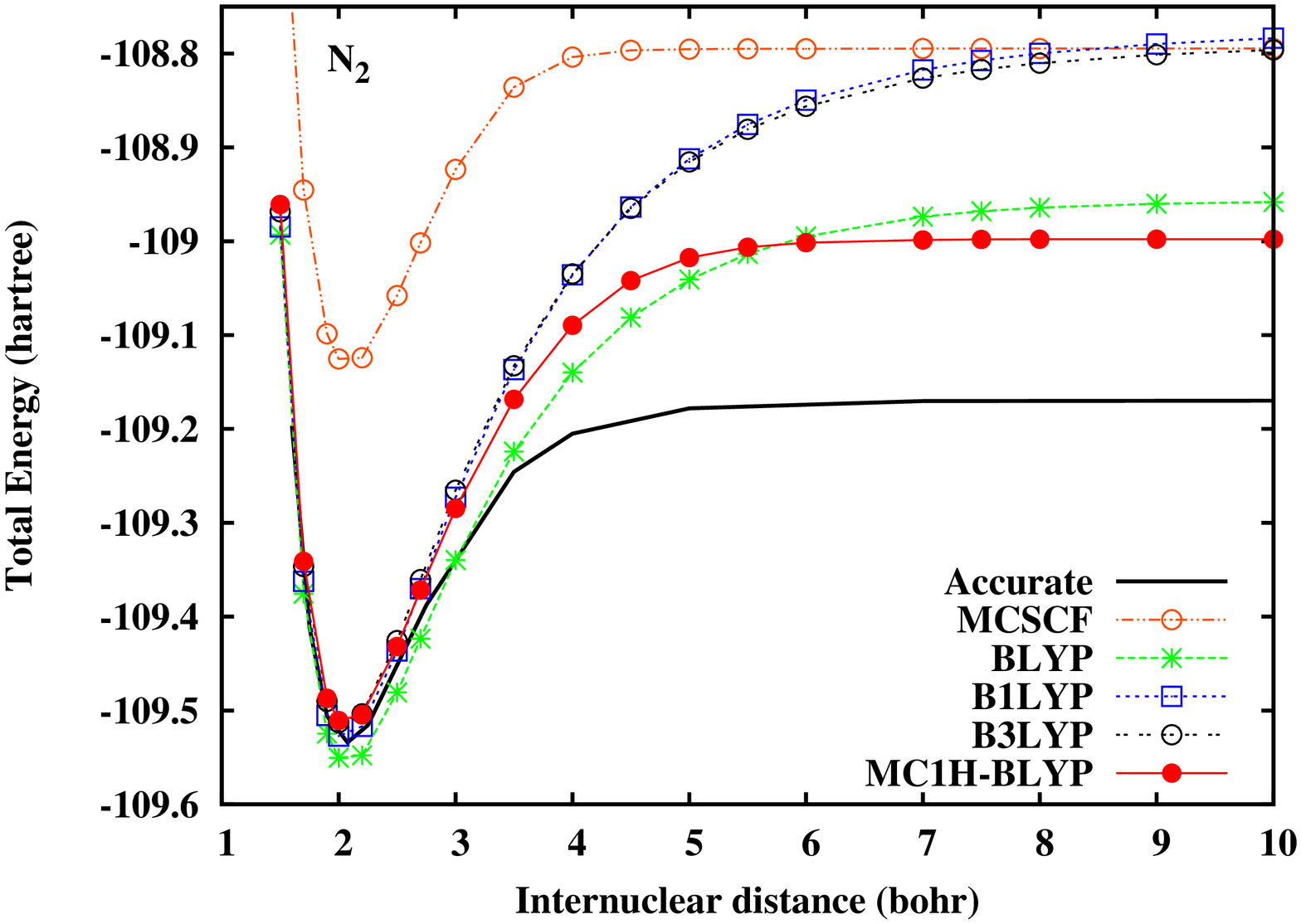}
\includegraphics[scale=0.3,angle=-90]{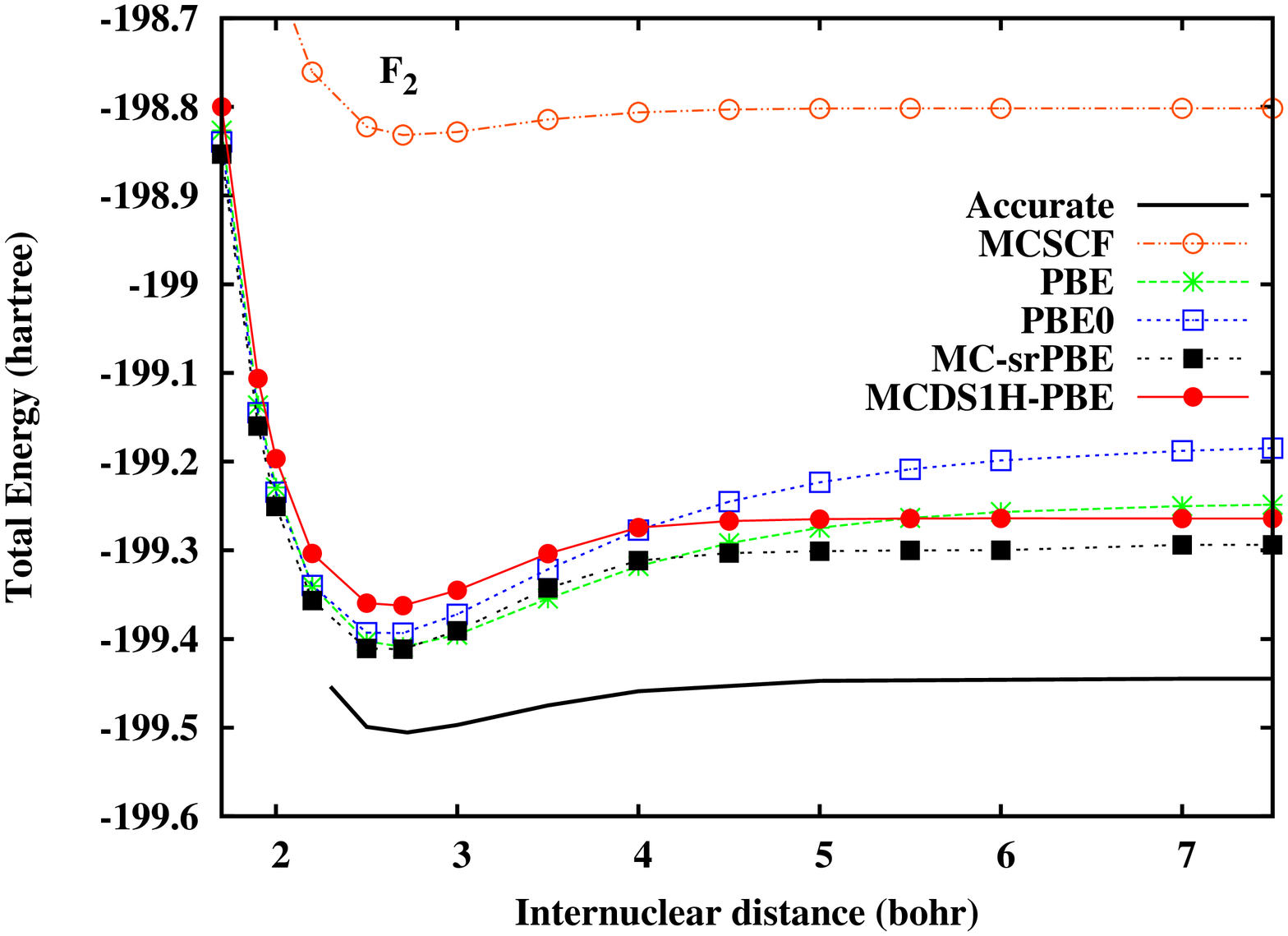}
\includegraphics[scale=0.3,angle=-90]{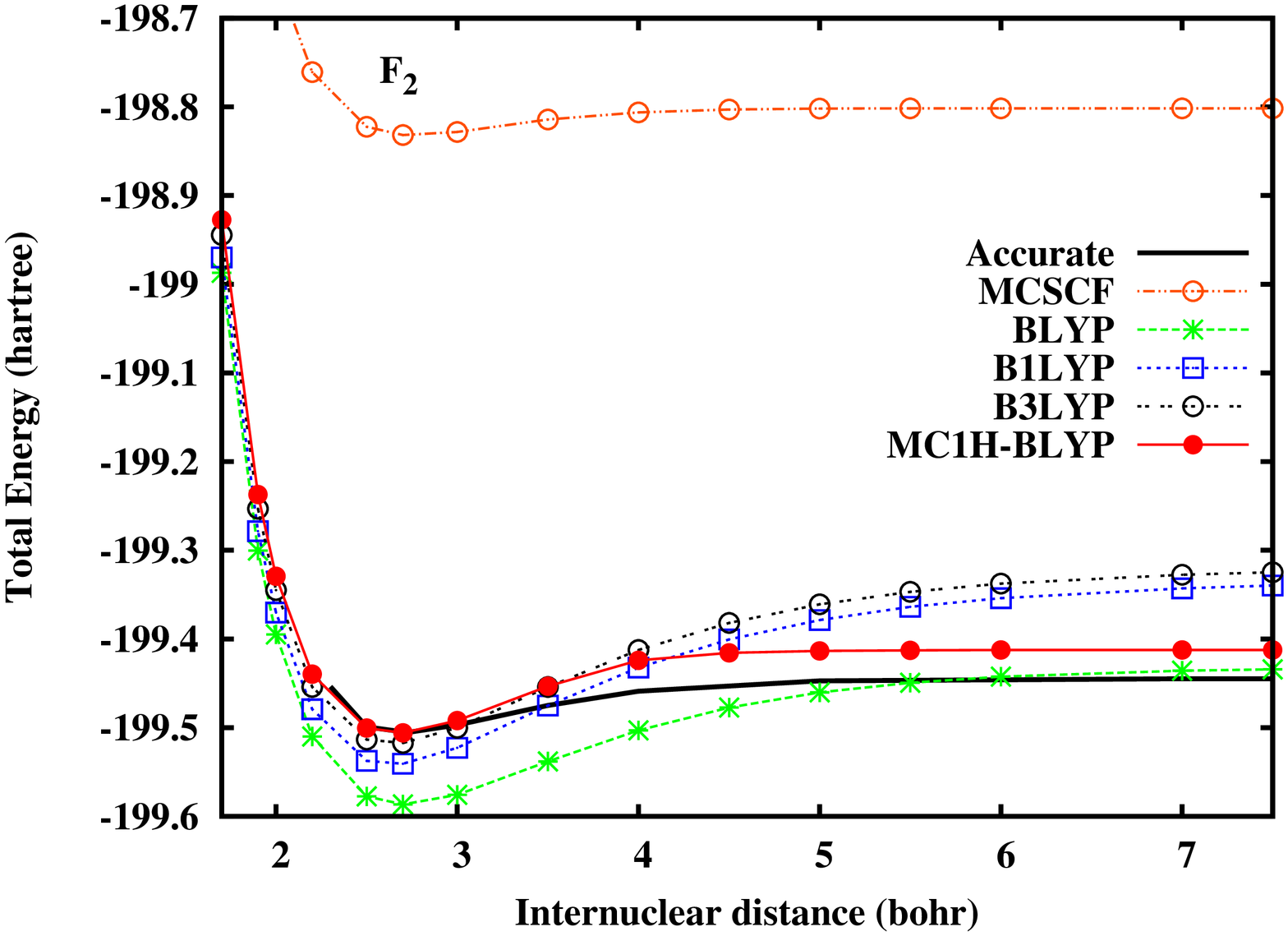}
\caption{Potential energy curves of Li$_2$, C$_2$, N$_2$, and F$_2$ calculated with MCSCF and several methods based on the PBE (left) or BLYP (right) exchange-correlation density functionals. For the MCDS1H and MC1H multiconfigurational hybrid approximations, we use a value of $\l=0.25$. For the range-separated MC-srPBE multiconfigurational hybrid approximation, we use a value of the range-separation parameter of $\mu=0.40$ bohr$^{-1}$. For all multiconfigurational methods, we use a full-valence CAS wave function. The basis set is cc-pVTZ basis. The accurate curves are from Ref.~\onlinecite{LieCle-JCP-74b}.}
\label{fig:diatomics}
\end{figure*}

The reaction energies of the formation of the cycloadducts are overestimated in MCSCF and underestimated in MRMP2 (by about 15 kcal/mol). Zhao \textit{et al.} observed that even using a large (14,14) active space does not improve the MRMP2 reaction energy~\cite{ZhaTisGouLiLutPieTru-JPCA-09}. All the hybrid methods give more reasonable reaction energies. In particular, MC1H-BLYP gives reaction energies within less than 3 kcal/mol of the best estimates.

If we accept to look more closely to the MAE values in spite of the limited statistics, we see that MC1H-BLYP gives overall the smallest MAE with 1.30 kcal/mol. The multiconfigurational hybrid MCDS1H-PBE involving the PBE exchange-correlation functional gives a larger MAE of 3.11 kcal/mol, but turns out to perform better on average than the range-separated MC-srPBE method which gives a MAE of 4.27 kcal/mol. Neglecting density scaling in the correlation functional of multiconfigurational hybrids seems slightly favorable for BLYP and slightly unfavorable for PBE. This is in line with what was found for double-hybrid approximations~\cite{ShaTouSav-JCP-11}, although the MP2 correlation part made it much more sensitive to the neglect of the density scaling. Even if the effect of neglecting density scaling is systematic in giving more negative complement correlation energies $\bar{E}^{\l}_c[n]$ for all density functionals, its effect on the MAE depends more on fortuitous compensation of errors for the approximate functional used.

\subsection{Dissociation of diatomic molecules}

We now turn to the calculation of potential energy curves of diatomic molecules. This is a harder problem since static correlation effects are dominant at dissociation. For the multiconfigurational hybrids, we report here only the curves of MCDS1H for PBE and of MC1H for BLYP according to the results of Section~\ref{sec:O3ADD6}, but the differences between the curves of MCDS1H and MC1H for both BLYP and PBE are in fact very small for these diatomic molecules.

Figure~\ref{fig:h2} shows the potential energy curve of H$_2$ calculated by hybrid approximations using the PBE and BLYP density functionals. Around the equilibrium internuclear distance, all DFT-based method, including MCDS1H-PBE and MC1H-BLYP, are accurate, which means that they properly describe dynamic correlation. At large distances, the single-hybrid approximations (PBE0, B1LYP, and B3LYP), which include a fraction of HF exchange energy, give less accurate potential energy curves than non-hybrid KS calculations (PBE and BLYP). By inclusion of a fraction of exact static correlation energy, the multiconfigurational hybrids (MCDS1H-PBE, MC-srPBE, and MC1H-BLYP) correct this behavior and give potential energy curves that correctly saturate beyond a distance of about $5$ bohr, as the MCSCF curve does. This point is explained by a detailed analysis of the asymptotic expansion of the potential energy curves in a minimal basis in Appendix~\ref{app:asymp}. However, the MCDS1H-PBE, MC-srPBE, and MC1H-BLYP methods still display significant errors on the energy of the separated atoms due to the density-functional approximations. Indeed, as in restricted KS calculations, the density functionals used in the multiconfigurational hybrids depend only of the total density and do not give accurate energies in the limit of separated atoms of open-shell character. In an unrestricted KS calculation, the energy at dissociation can be improved by breaking the spin symmetry and therefore introducing a fictitious spin density which helps to describe the separated atoms. In our present implementation of the multiconfigurational hybrids, the spin symmetry is imposed on the MCSCF wave function so that there is no fictitious spin density to be used in the density functionals.

Figure~\ref{fig:diatomics} shows the potential energy curves of Li$_2$, C$_2$, N$_2$, and F$_2$. The results are similar than for H$_2$. Around the equilibrium distance, the MCDS1H-PBE and MC1H-BLYP curves are similar to the standard hybrid or non-hybrid KS calculations. At large distances, the single-hybrid approximations give a spurious increase of the energy, whereas the MCDS1H-PBE and MC1H-BLYP curves correctly saturate. The MC1H-BLYP approximation gives good total energies, but again a significant error remains at dissociation, especially for N$_2$. The DS1DH-PBE and MC-srPBE approximation gives curves of very similar shape.

\section{Conclusions}

We have presented a multiconfigurational hybrid density-functional theory which rigorously combines MCSCF and DFT based on a linear decomposition of the electron-electron interaction. It is straightforward extension of the usual hybrid approximations by essentially adding a fraction $\l$ of exact static correlation in addition to the fraction $\l$ of exact exchange. Any existing approximate exchange-correlation density functional can be used in this scheme by using a simple scaling relation with $\l$. Test calculations on the cycloaddition reactions of ozone with ethylene or acetylene and the dissociation of diatomic molecules with the PBE and BLYP density functionals show that a good value of $\l$ is $0.25$, as in the usual hybrid approximations. 

Interestingly, the results seem to indicate that the present approach based on a simple linear decomposition of the electron-electron interaction is at least as good as the range-separated multiconfigurational hybrid method of Ref.~\onlinecite{PedJen-JJJ-XX,FroTouJen-JCP-07,FroReaWahWahJen-JCP-09} for including static correlation in DFT, at least with the approximate density functionals used here. Of course, with better short-range density-functional approximations (in particular, we do not have a short-range version of the LYP correlation functional for comparison), the conclusion could be different. Also, one should note that hybrid approaches combining perturbation theory with DFT based on a linear decomposition of the interaction~\cite{ShaTouSav-JCP-11} do not have the advantages of the range-separated hybrid approaches for fast basis-size convergence and explicit inclusion of long-range van der Waals interactions. For MCSCF, however, basis set convergence is not so much an issue.

The present results suggest that the proposed multiconfigurational hybrid approximations can improve over usual DFT approximations for situations with strong static correlation effects. It remains however to assess the performance of this multiconfigurational hybrid method on a larger variety of systems. Future work includes adding the dependence on the spin density in the functionals to be able to properly handle open-shell systems, and possibly other additional variables such as the on-top pair density as an alternative to the spin density for improving the accuracy of closed-shell systems~\cite{PerSavBur-PRA-95}.

\section*{Acknowledgments}
We thank E. Fromager (Strasbourg, France) for discussions. K.S. thanks Professors I. Othman and M. K. Sabra (Atomic Energy Commission of Syria) for their support. This work was part of the French-Danish collaboration project ``WADEMECOM.dk'' supported by the French government.

\appendix
\section{Scaling relations for the derivatives of the density-scaled correlation functional}
\label{app:scaling}

We give the scaling relations for the density-scaled correlation functional $E_c^\l[n]=\l^2 E_c[n_{1/\l}]$ and its derivatives in the case of generalized-gradient approximations (GGA). Starting from a standard GGA density functional written as
\begin{eqnarray}
E_{c,\GGA}[n] &=& \int e_{c} \left(n(\b{r}),\left|\b{\n} n(\b{r})\right|\right) d\b{r},
\end{eqnarray}
where $\left|\b{\n} n(\b{r})\right|$ is the norm of the density gradient, the corresponding scaled functional is 
\begin{eqnarray}
E_{c,\GGA}^\l[n] &=& \int e_{c}^\l \left(n(\b{r}),\left|\b{\n} n(\b{r})\right|\right) d\b{r},
\end{eqnarray}
where the energy density is obtained by scaling relation (see Ref.~\onlinecite{ShaTouSav-JCP-11})
\begin{eqnarray}
e_{c}^\l \left(n(\b{r}),\left|\b{\n} n(\b{r})\right|\right) = \l^5 e_{c} \left(\frac{n(\b{r})}{\l^3},\frac{\left|\b{\n} n(\b{r})\right|}{\l^4}\right).
\end{eqnarray}
The first-order derivatives of the energy density are
\begin{eqnarray}
\frac{\partial e_c^{\l}}{\partial n}\left( n(\b{r}), \left|\b{\n} n(\b{r})\right|\right) = \l^2 \frac{\partial e_c}{\partial n} \left(\frac{n(\b{r})}{\l^3},\frac{\left|\b{\n} n(\b{r})\right|}{\l^4} \right),
\label{Firstder}
\end{eqnarray}
and
\begin{eqnarray}
\frac{\partial e_c^{\l}}{\partial \left|\b{\n} n \right|}\left( n(\b{r}), \left|\b{\n} n(\b{r})\right|\right)  = \l \frac{\partial e_c}{\partial \left|\b{\n} n \right|} \left(\frac{n(\b{r})}{\l^3},\frac{\left|\b{\n} n(\b{r})\right|}{\l^4} \right).
\nonumber\\
\end{eqnarray}
The second-order derivatives are 
\begin{eqnarray}
\frac{\partial^{2} e_c^{\l}}{\partial n^{2}} \left( n(\b{r}), \left|\b{\n} n(\b{r})\right|\right) = \frac{1}{\l} \frac{\partial^{2} e_c}{\partial n^{2}} \left(\frac{n(\b{r})}{\l^3},\frac{\left|\b{\n} n(\b{r})\right|}{\l^4} \right),
\end{eqnarray}
\begin{eqnarray}
\frac{\partial^{2} e_c^{\l}}{\partial \left|\b{\n} n \right|^{2}} \left( n(\b{r}), \left|\b{\n} n(\b{r})\right|\right) = \;\;\;\;\;\;\;\;\;\;\;\;
\nonumber\\
\frac{1}{\l^{3}} \frac{\partial^2 e_c}{\partial \left|\b{\n} n \right|^2} \left(\frac{n(\b{r})}{\l^3},\frac{\left|\b{\n}n(\b{r})\right|}{\l^4} \right),
\end{eqnarray}
and
\begin{eqnarray}
\frac{\partial^{2} e_c^{\l}}{\partial n \partial \left|\b{\n} n \right|} \left( n(\b{r}), \left|\b{\n} n(\b{r})\right|\right) =\;\;\;\;\;\;\;\;\;\;\;\;
\nonumber\\
\frac{1}{\l^{2}} \frac{\partial^{2} e_c}{\partial n \partial \left|\b{\n} n \right| } \left(\frac{n(\b{r})}{\l^3},\frac{\left|\b{\n} n(\b{r})\right|}{\l^4} \right).
\end{eqnarray}

\section{Asymptotic expansion of the potential energy curve of H$_2$}
\label{app:asymp}

We consider the H$_2$ molecule in a Slater minimal basis, with a basis function $a$ localized on the left atom and a basis function $b$ localized on the right atom, both basis functions being identical with exponent $\zeta=1$. In the large internuclear distance $R$ limit, the two molecular orbitals are $1 = (a+b)/\sqrt{2}$ and $2 = (a-b)/\sqrt{2}$. The total restricted Hartree-Fock (RHF) energy writes
\begin{equation}
E^{\text{RHF}} = 2 h_{11} + J_{11} + \frac{1}{R},
\end{equation}
where $h_{11}= t_{11}+v_{11}$ is the sum of the kinetic integral $t_{11} = (1|\hat{t}|1)$ and the nuclei-electron integral $v_{11}=(1|\hat{v}_{ne}|1)$, and $J_{11}=(11|11)$ is the Coulomb two-electron integral, and $1/R$ is the nuclear repulsion energy. By expanding the molecular orbital $1$ into the localized functions $a$ and $b$, and using the symmetry between $a$ and $b$, it is easy to find the large $R$ behavior of all these terms:
\begin{eqnarray}
t_{11} = (a|\hat{t}|a) + (a|\hat{t}|b) = \frac{1}{2} +  O(e^{-R}),
\end{eqnarray}
and
\begin{eqnarray}
v_{11} = (a|\hat{v}_{ne}|a) + (a|\hat{v}_{ne}|b) = -1 -\frac{1}{R} +  O(e^{-R}),
\end{eqnarray}
and
\begin{eqnarray}
J_{11} &=& \frac{(aa|aa)}{2} + \frac{(aa|bb)}{2} + 2 (aa|ab) + (ab|ab) 
\nonumber\\
&=& \frac{5}{16} + \frac{1}{2R} + O(e^{-R}),
\end{eqnarray}
where $O(e^{-R})$ stands for exponentially decaying terms in $R$. For the values of the integrals, see Ref.~\onlinecite{DewKel-JCE-71}. Adding all the pieces together, it leads to the following asymptotic expansion of the total RHF energy
\begin{equation}
E^{\text{RHF}} = -\frac{11}{16} -\frac{1}{2R} + O(e^{-R}).
\label{EHFasymp}
\end{equation}
At dissociation, the RHF wave function contains 50\% of the incorrect ionic contribution H$^+$...H$^-$, which is responsible for too high an energy and for the spurious electrostatic attraction term $-1/2R$.

The full configuration interaction (FCI) correlation energy in this basis is found by diagonalizing the $2\times2$ Hamiltonian matrix, leading to
\begin{equation}
E_{c}^{\text{FCI}} = \frac{1}{2} \left(E_2 -E^{\text{RHF}} -\sqrt{(E_2-E^{\text{RHF}})^2 + 4 K_{12}^2} \right),
\end{equation}
where $E_2 = 2 h_{22} + J_{22} + 1/R$ is the energy of the double-excited determinant, and $K_{12} = (12|12)$ is the exchange two-electron integral. The asymptotic behavior of $E_2$ is exactly the same as the one of $E^{\text{RHF}}$, so that $E_2-E^{\text{RHF}}$ vanishes exponentially when $R\to\infty$ and the asymptotic behavior of $E_c^{\text{FCI}}$ is determined by $K_{12}$ only: $E_{c}^{\text{FCI}} = -K_{12} + O(e^{-R})$. The asymptotic behavior of $K_{12}$ is
\begin{eqnarray}
K_{12} &=& \frac{(aa|aa)}{2} - \frac{(aa|bb)}{2} = \frac{5}{16} - \frac{1}{2R} + O(e^{-R}),
\nonumber\\
\end{eqnarray}
giving for the correlation energy
\begin{equation}
E_{c}^{\text{FCI}} = -\frac{5}{16} + \frac{1}{2R} + O(e^{-R}).
\label{Ecasymp}
\end{equation}
Adding the asymptotic expansions of Eqs.~(\ref{EHFasymp}) and~(\ref{Ecasymp}) gives the asymptotic expansion of the total FCI energy in this basis
\begin{equation}
E^\text{FCI} = -1 + O(e^{-R}),
\end{equation}
which implies that the FCI potential energy curve saturates quickly at large internuclear distance.

In restricted Kohn-Sham (RKS) density-functional theory, the total energy writes
\begin{equation}
E^{\text{RKS}} = 2 h_{11} + 2J_{11} + E_{x} + E_c + \frac{1}{R},
\end{equation}
where $E_{x}$ and $E_c$ are the exchange and correlation energies. At large $R$, it behaves as
\begin{equation}
E^{\text{RKS}} = -\frac{3}{8} + E_{x} + E_c+ O(e^{-R}).
\label{EKSasymp}
\end{equation}
With local or semilocal density-functional approximations, $E_{x}$ and $E_c$ go exponentially to constants when $R\to\infty$, so that the asymptotic expansion of $E^{\text{RKS}}$ does not contain a spurious term in $1/R$. 

Single-hybrid approximations introduces a fraction $\l$ of RHF exchange which have the following asymptotic expansion
\begin{equation}
E_x^{\text{RHF}}=-J_{11} = -\frac{5}{16} -\frac{1}{2R} + O(e^{-R}),
\end{equation}
and therefore introduce a wrong $-\l/2R$ term in the total energy,
\begin{equation}
E^{\text{hybrid}} = -\frac{3}{8} - \frac{5\l}{16} +(1-\l)E_{x} + E_c -\frac{\l}{2R} + O(e^{-R}).
\label{}
\end{equation}
Single-hybrid approximations thus deteriorate the large $R$ behavior of local or semilocal density-functional approximations (see Fig.~\ref{fig:h2}). The multiconfigurational hybrid approximations introduced in this work correct this behavior by adding a fraction of the FCI correlation energy which, in the limit of large $R$, is just $\l E_{c}^{\text{FCI}}$, the linearity in $\l$ being a signature of static correlation. For example, the MC1H approximation has the following asymptotic expansion
\begin{equation}
E^{\text{MC1H}} = -\frac{3}{8} - \frac{5\l}{8} +(1-\l)E_{x} + (1-\l^2)E_c + O(e^{-R}),
\label{}
\end{equation}
with no longer any spurious $1/R$ term, and thus improves the large $R$ behavior (see Fig.~\ref{fig:h2}). It is a typical example where exact exchange and static correlation must be considered together.

For range-separated density-functional theory, the situation is similar. Range-separated single-hybrid approximations~\cite{IikTsuYanHir-JCP-01,YanTewHan-CPL-04,AngGerSavTou-PRA-05,GerAng-CPL-05a,VydScu-JCP-06} include some long-range RHF exchange and their asymptotic expansions display a wrong $-1/2R$ term, just as RHF. Their behavior for large $R$ is in fact worse than that of usual single-hybrid approximations since the $-1/2R$ term is not weighted by $\l$. However, the range-separated CI-srDFT~\cite{LeiStoWerSav-CPL-97,PolSavLeiSto-JCP-02} or MC-srDFT~\cite{PedJen-JJJ-XX,FroTouJen-JCP-07,FroReaWahWahJen-JCP-09} methods add some exact long-range correlation energy which removes this wrong $-1/2R$ term. 

Note that other forms of single-hybrid approximations which do not use RHF exchange at long range~\cite{HeyScuErn-JCP-03,HenIzmScuSav-JCP-07} allow one to avoid a wrong $-1/2R$ term in the large $R$ limit. Symmetry breaking is another way to avoid a wrong asymptotic $-1/2R$ term since the unrestricted Hartree-Fock exchange energy does not contain such a term.

The fact that local or semilocal approximations for $E_{x}$ and $E_c$ do not introduce $1/R$ terms is in agreement with the usual conviction that approximate GGA exchange functionals not only represent exchange but also static correlation, while approximate GGA correlation functionals represent dynamic correlation only~\cite{GriSchBae-JCP-97}. 



\begin{thebibliography}{0}
\expandafter\ifx\csname natexlab\endcsname\relax\def\natexlab#1{#1}\fi
\expandafter\ifx\csname bibnamefont\endcsname\relax
  \def\bibnamefont#1{#1}\fi
\expandafter\ifx\csname bibfnamefont\endcsname\relax
  \def\bibfnamefont#1{#1}\fi
\expandafter\ifx\csname citenamefont\endcsname\relax
  \def\citenamefont#1{#1}\fi
\expandafter\ifx\csname url\endcsname\relax
  \def\url#1{\texttt{#1}}\fi
\expandafter\ifx\csname urlprefix\endcsname\relax\def\urlprefix{URL }\fi
\providecommand{\bibinfo}[2]{#2}
\providecommand{\eprint}[2][]{\url{#2}}

\end{thebibliography}


\begin{thebibliography}{128}
\expandafter\ifx\csname natexlab\endcsname\relax\def\natexlab#1{#1}\fi
\expandafter\ifx\csname bibnamefont\endcsname\relax
  \def\bibnamefont#1{#1}\fi
\expandafter\ifx\csname bibfnamefont\endcsname\relax
  \def\bibfnamefont#1{#1}\fi
\expandafter\ifx\csname citenamefont\endcsname\relax
  \def\citenamefont#1{#1}\fi
\expandafter\ifx\csname url\endcsname\relax
  \def\url#1{\texttt{#1}}\fi
\expandafter\ifx\csname urlprefix\endcsname\relax\def\urlprefix{URL }\fi
\providecommand{\bibinfo}[2]{#2}
\providecommand{\eprint}[2][]{\url{#2}}

\bibitem[{\citenamefont{Hohenberg and Kohn}(1964)}]{HohKoh-PR-64}
\bibinfo{author}{\bibfnamefont{P.}~\bibnamefont{Hohenberg}} \bibnamefont{and}
  \bibinfo{author}{\bibfnamefont{W.}~\bibnamefont{Kohn}},
  \bibinfo{journal}{Phys. Rev.} \textbf{\bibinfo{volume}{{136}}},
  \bibinfo{pages}{B 864} (\bibinfo{year}{1964}).

\bibitem[{\citenamefont{Kohn and Sham}(1965)}]{KohSha-PR-65}
\bibinfo{author}{\bibfnamefont{W.}~\bibnamefont{Kohn}} \bibnamefont{and}
  \bibinfo{author}{\bibfnamefont{L.~J.} \bibnamefont{Sham}},
  \bibinfo{journal}{Phys. Rev.} \textbf{\bibinfo{volume}{140}},
  \bibinfo{pages}{A1133} (\bibinfo{year}{1965}).

\bibitem[{\citenamefont{Koch and Holthausen}(2001)}]{KocHol-BOOK-01}
\bibinfo{author}{\bibfnamefont{W.}~\bibnamefont{Koch}} \bibnamefont{and}
  \bibinfo{author}{\bibfnamefont{M.~C.} \bibnamefont{Holthausen}},
  \emph{\bibinfo{title}{A Chemist's Guide To Density Functional Theory}}
  (\bibinfo{publisher}{Wiley-VCH}, \bibinfo{address}{New York},
  \bibinfo{year}{2001}).

\bibitem[{\citenamefont{Gritsenko et~al.}(1997)\citenamefont{Gritsenko,
  Schipper, and Baerends}}]{GriSchBae-JCP-97}
\bibinfo{author}{\bibfnamefont{O.~V.} \bibnamefont{Gritsenko}},
  \bibinfo{author}{\bibfnamefont{P.~R.~T.} \bibnamefont{Schipper}},
  \bibnamefont{and} \bibinfo{author}{\bibfnamefont{E.~J.}
  \bibnamefont{Baerends}}, \bibinfo{journal}{J. Chem. Phys.}
  \textbf{\bibinfo{volume}{107}}, \bibinfo{pages}{5007} (\bibinfo{year}{1997}).

\bibitem[{\citenamefont{Becke}(2003)}]{Bec-JCP-03}
\bibinfo{author}{\bibfnamefont{A.~D.} \bibnamefont{Becke}},
  \bibinfo{journal}{J. Chem. Phys.} \textbf{\bibinfo{volume}{119}},
  \bibinfo{pages}{2972} (\bibinfo{year}{2003}).

\bibitem[{\citenamefont{Cohen et~al.}(2008)\citenamefont{Cohen, Mori-S\'anchez,
  and Yang}}]{CohMorYan-SCI-08}
\bibinfo{author}{\bibfnamefont{A.~J.} \bibnamefont{Cohen}},
  \bibinfo{author}{\bibfnamefont{P.}~\bibnamefont{Mori-S\'anchez}},
  \bibnamefont{and} \bibinfo{author}{\bibfnamefont{W.}~\bibnamefont{Yang}},
  \bibinfo{journal}{Science} \textbf{\bibinfo{volume}{321}},
  \bibinfo{pages}{792} (\bibinfo{year}{2008}).

\bibitem[{\citenamefont{Schultz et~al.}(2005)\citenamefont{Schultz, Zhao, and
  Truhlar}}]{SchZhaTru-JPCA-05}
\bibinfo{author}{\bibfnamefont{N.~E.} \bibnamefont{Schultz}},
  \bibinfo{author}{\bibfnamefont{Y.}~\bibnamefont{Zhao}}, \bibnamefont{and}
  \bibinfo{author}{\bibfnamefont{D.~G.} \bibnamefont{Truhlar}},
  \bibinfo{journal}{J. Phys. Chem. A} \textbf{\bibinfo{volume}{109}},
  \bibinfo{pages}{11127} (\bibinfo{year}{2005}).

\bibitem[{\citenamefont{Cremer}(2001)}]{Cre-MP-01}
\bibinfo{author}{\bibfnamefont{D.}~\bibnamefont{Cremer}},
  \bibinfo{journal}{Mol. Phys.} \textbf{\bibinfo{volume}{99}},
  \bibinfo{pages}{1899} (\bibinfo{year}{2001}).

\bibitem[{\citenamefont{Slater et~al.}(1969)\citenamefont{Slater, Mann, Wilson,
  and Wood}}]{SlaManWilWoo-PR-69}
\bibinfo{author}{\bibfnamefont{J.}~\bibnamefont{Slater}},
  \bibinfo{author}{\bibfnamefont{J.}~\bibnamefont{Mann}},
  \bibinfo{author}{\bibfnamefont{T.}~\bibnamefont{Wilson}}, \bibnamefont{and}
  \bibinfo{author}{\bibfnamefont{J.}~\bibnamefont{Wood}},
  \bibinfo{journal}{Phys. Rev.} \textbf{\bibinfo{volume}{184}},
  \bibinfo{pages}{672} (\bibinfo{year}{1969}).

\bibitem[{\citenamefont{Levy}(1982)}]{Lev-PRA-82}
\bibinfo{author}{\bibfnamefont{M.}~\bibnamefont{Levy}}, \bibinfo{journal}{Phys.
  Rev. A} \textbf{\bibinfo{volume}{26}}, \bibinfo{pages}{1200}
  (\bibinfo{year}{1982}).

\bibitem[{\citenamefont{Lieb}(1983)}]{Lie-IJQC-83}
\bibinfo{author}{\bibfnamefont{E.~H.} \bibnamefont{Lieb}},
  \bibinfo{journal}{Int. J. Quantum. Chem.} \textbf{\bibinfo{volume}{{24}}},
  \bibinfo{pages}{24} (\bibinfo{year}{1983}).

\bibitem[{\citenamefont{Dunlap and Mei}(1983)}]{DunMei-JCP-83}
\bibinfo{author}{\bibfnamefont{B.~I.} \bibnamefont{Dunlap}} \bibnamefont{and}
  \bibinfo{author}{\bibfnamefont{W.~N.} \bibnamefont{Mei}},
  \bibinfo{journal}{J. Chem. Phys.} \textbf{\bibinfo{volume}{78}},
  \bibinfo{pages}{4997} (\bibinfo{year}{1983}).

\bibitem[{\citenamefont{Wang and Schwarz}(1996)}]{WanSch-JCP-96}
\bibinfo{author}{\bibfnamefont{S.~G.} \bibnamefont{Wang}} \bibnamefont{and}
  \bibinfo{author}{\bibfnamefont{W.~H.~E.} \bibnamefont{Schwarz}},
  \bibinfo{journal}{J. Chem. Phys.} \textbf{\bibinfo{volume}{{105}}},
  \bibinfo{pages}{4641} (\bibinfo{year}{1996}).

\bibitem[{\citenamefont{Schipper et~al.}(1998)\citenamefont{Schipper,
  Gritsenko, and Baerends}}]{SchGriBae-TCA-98}
\bibinfo{author}{\bibfnamefont{P.~R.~T.} \bibnamefont{Schipper}},
  \bibinfo{author}{\bibfnamefont{O.~V.} \bibnamefont{Gritsenko}},
  \bibnamefont{and} \bibinfo{author}{\bibfnamefont{E.~J.}
  \bibnamefont{Baerends}}, \bibinfo{journal}{Theor. Chem. Acc.}
  \textbf{\bibinfo{volume}{99}}, \bibinfo{pages}{329} (\bibinfo{year}{1998}).

\bibitem[{\citenamefont{Goddard and Orlova}(1999)}]{GodOrl-JCP-99}
\bibinfo{author}{\bibfnamefont{J.~D.} \bibnamefont{Goddard}} \bibnamefont{and}
  \bibinfo{author}{\bibfnamefont{G.}~\bibnamefont{Orlova}},
  \bibinfo{journal}{J. Chem. Phys.} \textbf{\bibinfo{volume}{111}},
  \bibinfo{pages}{7705} (\bibinfo{year}{1999}).

\bibitem[{\citenamefont{Filatov and Shaik}(1999{\natexlab{a}})}]{FilSha-CPL-99}
\bibinfo{author}{\bibfnamefont{M.}~\bibnamefont{Filatov}} \bibnamefont{and}
  \bibinfo{author}{\bibfnamefont{S.}~\bibnamefont{Shaik}},
  \bibinfo{journal}{Chem. Phys. Lett.} \textbf{\bibinfo{volume}{304}},
  \bibinfo{pages}{429} (\bibinfo{year}{1999}{\natexlab{a}}).

\bibitem[{\citenamefont{Filatov and
  Shaik}(1999{\natexlab{b}})}]{FilSha-JPCA-99}
\bibinfo{author}{\bibfnamefont{M.}~\bibnamefont{Filatov}} \bibnamefont{and}
  \bibinfo{author}{\bibfnamefont{S.}~\bibnamefont{Shaik}}, \bibinfo{journal}{J.
  Phys. Chem. A} \textbf{\bibinfo{volume}{103}}, \bibinfo{pages}{8885}
  (\bibinfo{year}{1999}{\natexlab{b}}).

\bibitem[{\citenamefont{Filatov and Shaik}(1999{\natexlab{c}})}]{FilSha-CPL-00}
\bibinfo{author}{\bibfnamefont{M.}~\bibnamefont{Filatov}} \bibnamefont{and}
  \bibinfo{author}{\bibfnamefont{S.}~\bibnamefont{Shaik}},
  \bibinfo{journal}{Chem. Phys. Lett.} \textbf{\bibinfo{volume}{332}},
  \bibinfo{pages}{409} (\bibinfo{year}{1999}{\natexlab{c}}).

\bibitem[{\citenamefont{Filatov et~al.}(2000)\citenamefont{Filatov, Shaik,
  Woeller, Grimme, and Peyerimhoff}}]{FilShaWoeGriPey-CPL-00}
\bibinfo{author}{\bibfnamefont{M.}~\bibnamefont{Filatov}},
  \bibinfo{author}{\bibfnamefont{S.}~\bibnamefont{Shaik}},
  \bibinfo{author}{\bibfnamefont{M.}~\bibnamefont{Woeller}},
  \bibinfo{author}{\bibfnamefont{S.}~\bibnamefont{Grimme}}, \bibnamefont{and}
  \bibinfo{author}{\bibfnamefont{S.}~\bibnamefont{Peyerimhoff}},
  \bibinfo{journal}{Chem. Phys. Lett.} \textbf{\bibinfo{volume}{316}},
  \bibinfo{pages}{135} (\bibinfo{year}{2000}).

\bibitem[{\citenamefont{Filatov and Shaik}(2000)}]{FilSha-JPCA-00}
\bibinfo{author}{\bibfnamefont{M.}~\bibnamefont{Filatov}} \bibnamefont{and}
  \bibinfo{author}{\bibfnamefont{S.}~\bibnamefont{Shaik}}, \bibinfo{journal}{J.
  Phys. Chem. A} \textbf{\bibinfo{volume}{104}}, \bibinfo{pages}{6628}
  (\bibinfo{year}{2000}).

\bibitem[{\citenamefont{Ullrich and Kohn}(2001)}]{UllKoh-PRL-01}
\bibinfo{author}{\bibfnamefont{C.~A.} \bibnamefont{Ullrich}} \bibnamefont{and}
  \bibinfo{author}{\bibfnamefont{W.}~\bibnamefont{Kohn}},
  \bibinfo{journal}{Phys. Rev. Lett.} \textbf{\bibinfo{volume}{87}},
  \bibinfo{pages}{093001} (\bibinfo{year}{2001}).

\bibitem[{\citenamefont{R.~Takeda}(2003)}]{TakYamYam-IJQC-03}
\bibinfo{author}{\bibfnamefont{K.~Y.} \bibnamefont{R.~Takeda},
  \bibfnamefont{S.~Yamanaka}}, \bibinfo{journal}{Int. J. Quantum. Chem.}
  \textbf{\bibinfo{volume}{{93}}}, \bibinfo{pages}{317} (\bibinfo{year}{2003}).

\bibitem[{\citenamefont{Grimme and Waletzke}(1999)}]{GriWal-JCP-99}
\bibinfo{author}{\bibfnamefont{S.}~\bibnamefont{Grimme}} \bibnamefont{and}
  \bibinfo{author}{\bibfnamefont{M.}~\bibnamefont{Waletzke}},
  \bibinfo{journal}{J. Chem. Phys.} \textbf{\bibinfo{volume}{111}},
  \bibinfo{pages}{5645} (\bibinfo{year}{1999}).

\bibitem[{\citenamefont{Beck et~al.}(2008)\citenamefont{Beck, Stahlberg,
  Burggraf, and Blaudeau}}]{BecStaBurBla-CP-08}
\bibinfo{author}{\bibfnamefont{E.~V.} \bibnamefont{Beck}},
  \bibinfo{author}{\bibfnamefont{E.~A.} \bibnamefont{Stahlberg}},
  \bibinfo{author}{\bibfnamefont{L.~W.} \bibnamefont{Burggraf}},
  \bibnamefont{and} \bibinfo{author}{\bibfnamefont{J.-P.}
  \bibnamefont{Blaudeau}}, \bibinfo{journal}{Chem. Phys.}
  \textbf{\bibinfo{volume}{349}}, \bibinfo{pages}{158} (\bibinfo{year}{2008}).

\bibitem[{\citenamefont{{Q. Wu, C.-L. Cheng, and T. Van
  Voorhis}}(2007)}]{WuCheVoo-JCP-07}
\bibinfo{author}{\bibnamefont{{Q. Wu, C.-L. Cheng, and T. Van Voorhis}}},
  \bibinfo{journal}{J. Chem. Phys.} \textbf{\bibinfo{volume}{127}},
  \bibinfo{pages}{164119} (\bibinfo{year}{2007}).

\bibitem[{\citenamefont{{Q. Wu, B. Kaduk, and T. Van
  Voorhis}}(2009)}]{WuKadVoo-JCP-09}
\bibinfo{author}{\bibnamefont{{Q. Wu, B. Kaduk, and T. Van Voorhis}}},
  \bibinfo{journal}{J. Chem. Phys.} \textbf{\bibinfo{volume}{130}},
  \bibinfo{pages}{034109} (\bibinfo{year}{2009}).

\bibitem[{\citenamefont{Lie and Clementi}(1974{\natexlab{a}})}]{LieCle-JCP-74a}
\bibinfo{author}{\bibfnamefont{G.~C.} \bibnamefont{Lie}} \bibnamefont{and}
  \bibinfo{author}{\bibfnamefont{E.}~\bibnamefont{Clementi}},
  \bibinfo{journal}{J. Chem. Phys.} \textbf{\bibinfo{volume}{60}},
  \bibinfo{pages}{1275} (\bibinfo{year}{1974}{\natexlab{a}}).

\bibitem[{\citenamefont{Lie and Clementi}(1974{\natexlab{b}})}]{LieCle-JCP-74b}
\bibinfo{author}{\bibfnamefont{G.~C.} \bibnamefont{Lie}} \bibnamefont{and}
  \bibinfo{author}{\bibfnamefont{E.}~\bibnamefont{Clementi}},
  \bibinfo{journal}{J. Chem. Phys.} \textbf{\bibinfo{volume}{60}},
  \bibinfo{pages}{1288} (\bibinfo{year}{1974}{\natexlab{b}}).

\bibitem[{\citenamefont{Colle and Salvetti}(1975)}]{ColSal-TCA-75}
\bibinfo{author}{\bibfnamefont{R.}~\bibnamefont{Colle}} \bibnamefont{and}
  \bibinfo{author}{\bibfnamefont{O.}~\bibnamefont{Salvetti}},
  \bibinfo{journal}{Theor. Chim. Acta} \textbf{\bibinfo{volume}{37}},
  \bibinfo{pages}{329} (\bibinfo{year}{1975}).

\bibitem[{\citenamefont{Colle and Salvetti}(1979)}]{ColSal-TCA-79}
\bibinfo{author}{\bibfnamefont{R.}~\bibnamefont{Colle}} \bibnamefont{and}
  \bibinfo{author}{\bibfnamefont{O.}~\bibnamefont{Salvetti}},
  \bibinfo{journal}{Theor. Chim. Acta} \textbf{\bibinfo{volume}{53}},
  \bibinfo{pages}{55} (\bibinfo{year}{1979}).

\bibitem[{\citenamefont{Colle and Salvetti}(1983)}]{ColSal-JCP-83}
\bibinfo{author}{\bibfnamefont{R.}~\bibnamefont{Colle}} \bibnamefont{and}
  \bibinfo{author}{\bibfnamefont{O.}~\bibnamefont{Salvetti}},
  \bibinfo{journal}{J. Chem. Phys.} \textbf{\bibinfo{volume}{79}},
  \bibinfo{pages}{1404} (\bibinfo{year}{1983}).

\bibitem[{\citenamefont{Colle and Salvetti}(1990)}]{ColSal-JCP-90}
\bibinfo{author}{\bibfnamefont{R.}~\bibnamefont{Colle}} \bibnamefont{and}
  \bibinfo{author}{\bibfnamefont{O.}~\bibnamefont{Salvetti}},
  \bibinfo{journal}{J. Chem. Phys.} \textbf{\bibinfo{volume}{93}}
  (\bibinfo{year}{1990}).

\bibitem[{\citenamefont{Savin}(1988)}]{Sav-IJQC-88}
\bibinfo{author}{\bibfnamefont{A.}~\bibnamefont{Savin}}, \bibinfo{journal}{Int.
  J. Quantum. Chem.} \textbf{\bibinfo{volume}{{22}}}, \bibinfo{pages}{59}
  (\bibinfo{year}{1988}).

\bibitem[{\citenamefont{Savin}(1989)}]{Sav-JCP-89}
\bibinfo{author}{\bibfnamefont{A.}~\bibnamefont{Savin}}, \bibinfo{journal}{J.
  chim. phys.} \textbf{\bibinfo{volume}{{86}}}, \bibinfo{pages}{757}
  (\bibinfo{year}{1989}).

\bibitem[{\citenamefont{Savin}(1991)}]{Sav-INC-91}
\bibinfo{author}{\bibfnamefont{A.}~\bibnamefont{Savin}}, in
  \emph{\bibinfo{booktitle}{Density functional methods in chemistry}}, edited
  by \bibinfo{editor}{\bibfnamefont{J.}~\bibnamefont{Labanowski}}
  \bibnamefont{and} \bibinfo{editor}{\bibfnamefont{J.}~\bibnamefont{Andzelm}}
  (\bibinfo{publisher}{Springer-Verlag}, \bibinfo{address}{New York},
  \bibinfo{year}{1991}), p. \bibinfo{pages}{213}.

\bibitem[{\citenamefont{Miehlich et~al.}(1997)\citenamefont{Miehlich, Stoll,
  and Savin}}]{MieStoSav-MP-97}
\bibinfo{author}{\bibfnamefont{B.}~\bibnamefont{Miehlich}},
  \bibinfo{author}{\bibfnamefont{H.}~\bibnamefont{Stoll}}, \bibnamefont{and}
  \bibinfo{author}{\bibfnamefont{A.}~\bibnamefont{Savin}},
  \bibinfo{journal}{Mol. Phys.} \textbf{\bibinfo{volume}{{91}}},
  \bibinfo{pages}{527} (\bibinfo{year}{1997}).

\bibitem[{\citenamefont{Gutl\'e and Savin}(2007)}]{GutSav-PRA-07}
\bibinfo{author}{\bibfnamefont{C.}~\bibnamefont{Gutl\'e}} \bibnamefont{and}
  \bibinfo{author}{\bibfnamefont{A.}~\bibnamefont{Savin}},
  \bibinfo{journal}{Phys. Rev. A} \textbf{\bibinfo{volume}{75}},
  \bibinfo{pages}{032519} (\bibinfo{year}{2007}).

\bibitem[{\citenamefont{Moscardo and San-Fabian}(1991)}]{MosSan-IJQC-91}
\bibinfo{author}{\bibfnamefont{F.}~\bibnamefont{Moscardo}} \bibnamefont{and}
  \bibinfo{author}{\bibfnamefont{E.}~\bibnamefont{San-Fabian}},
  \bibinfo{journal}{Int. J. Quantum. Chem.} \textbf{\bibinfo{volume}{{40}}},
  \bibinfo{pages}{23} (\bibinfo{year}{1991}).

\bibitem[{\citenamefont{Kraka}(1992)}]{Kra-CP-92}
\bibinfo{author}{\bibfnamefont{E.}~\bibnamefont{Kraka}},
  \bibinfo{journal}{Chem. Phys.} \textbf{\bibinfo{volume}{161}},
  \bibinfo{pages}{141} (\bibinfo{year}{1992}).

\bibitem[{\citenamefont{Kraka et~al.}(1992)\citenamefont{Kraka, Cremer, and
  Nordholm}}]{KraCreNor-INC-92}
\bibinfo{author}{\bibfnamefont{E.}~\bibnamefont{Kraka}},
  \bibinfo{author}{\bibfnamefont{D.}~\bibnamefont{Cremer}}, \bibnamefont{and}
  \bibinfo{author}{\bibfnamefont{S.}~\bibnamefont{Nordholm}}, in
  \emph{\bibinfo{booktitle}{Molecules in Natural Science and Biomedicine}},
  edited by \bibinfo{editor}{\bibfnamefont{Z.}~\bibnamefont{Maksic}}
  \bibnamefont{and}
  \bibinfo{editor}{\bibfnamefont{M.}~\bibnamefont{Eckert-Maksic}}
  (\bibinfo{publisher}{Ellis Horwood}, \bibinfo{address}{Chichester},
  \bibinfo{year}{1992}), p. \bibinfo{pages}{351}.

\bibitem[{\citenamefont{Wu and Shaik}(1999)}]{WuSha-CPL-99}
\bibinfo{author}{\bibfnamefont{W.}~\bibnamefont{Wu}} \bibnamefont{and}
  \bibinfo{author}{\bibfnamefont{S.}~\bibnamefont{Shaik}},
  \bibinfo{journal}{Chem. Phys. Lett.} \textbf{\bibinfo{volume}{301}},
  \bibinfo{pages}{37} (\bibinfo{year}{1999}).

\bibitem[{\citenamefont{Stoll}(2003)}]{Sto-CPL-03}
\bibinfo{author}{\bibfnamefont{H.}~\bibnamefont{Stoll}},
  \bibinfo{journal}{Chem. Phys. Lett.} \textbf{\bibinfo{volume}{{376}}},
  \bibinfo{pages}{141} (\bibinfo{year}{2003}).

\bibitem[{\citenamefont{Ying et~al.}(2012)\citenamefont{Ying, Su, Chen, Shaik,
  and Wu}}]{YinSuCheShaWu-JCTC-12}
\bibinfo{author}{\bibfnamefont{F.}~\bibnamefont{Ying}},
  \bibinfo{author}{\bibfnamefont{P.}~\bibnamefont{Su}},
  \bibinfo{author}{\bibfnamefont{Z.}~\bibnamefont{Chen}},
  \bibinfo{author}{\bibfnamefont{S.}~\bibnamefont{Shaik}}, \bibnamefont{and}
  \bibinfo{author}{\bibfnamefont{W.}~\bibnamefont{Wu}}, \bibinfo{journal}{J.
  Chem. Theory Comput.} \textbf{\bibinfo{volume}{8}}, \bibinfo{pages}{1608}
  (\bibinfo{year}{2012}).

\bibitem[{\citenamefont{Malcolm and McDouall}(1994)}]{MalMcd-JPC-94}
\bibinfo{author}{\bibfnamefont{N.~O.~J.} \bibnamefont{Malcolm}}
  \bibnamefont{and} \bibinfo{author}{\bibfnamefont{J.~J.~W.}
  \bibnamefont{McDouall}}, \bibinfo{journal}{J. Phys. Chem.}
  \textbf{\bibinfo{volume}{98}}, \bibinfo{pages}{12579} (\bibinfo{year}{1994}).

\bibitem[{\citenamefont{Malcolm and McDouall}(1996)}]{MalMcd-JPC-96}
\bibinfo{author}{\bibfnamefont{N.~O.~J.} \bibnamefont{Malcolm}}
  \bibnamefont{and} \bibinfo{author}{\bibfnamefont{J.~J.~W.}
  \bibnamefont{McDouall}}, \bibinfo{journal}{J. Phys. Chem.}
  \textbf{\bibinfo{volume}{100}}, \bibinfo{pages}{10131}
  (\bibinfo{year}{1996}).

\bibitem[{\citenamefont{Malcolm and McDouall}(1997)}]{MalMcd-JPCA-97}
\bibinfo{author}{\bibfnamefont{N.~O.~J.} \bibnamefont{Malcolm}}
  \bibnamefont{and} \bibinfo{author}{\bibfnamefont{J.~J.~W.}
  \bibnamefont{McDouall}}, \bibinfo{journal}{J. Phys. Chem. A}
  \textbf{\bibinfo{volume}{101}}, \bibinfo{pages}{8119} (\bibinfo{year}{1997}).

\bibitem[{\citenamefont{Malcolm and McDouall}(1998)}]{MalMcd-CPL-98}
\bibinfo{author}{\bibfnamefont{N.~O.~J.} \bibnamefont{Malcolm}}
  \bibnamefont{and} \bibinfo{author}{\bibfnamefont{J.~J.~W.}
  \bibnamefont{McDouall}}, \bibinfo{journal}{Chem. Phys. Lett.}
  \textbf{\bibinfo{volume}{282}}, \bibinfo{pages}{121} (\bibinfo{year}{1998}).

\bibitem[{\citenamefont{McDouall}(2003)}]{Mcd-MP-03}
\bibinfo{author}{\bibfnamefont{J.~J.~W.} \bibnamefont{McDouall}},
  \bibinfo{journal}{Mol. Phys.} \textbf{\bibinfo{volume}{101}},
  \bibinfo{pages}{361} (\bibinfo{year}{2003}).

\bibitem[{\citenamefont{Borowski et~al.}(1998)\citenamefont{Borowski, Jordan,
  Nichols, and Nachtigall}}]{BorJorNicNac-TCA-98}
\bibinfo{author}{\bibfnamefont{P.}~\bibnamefont{Borowski}},
  \bibinfo{author}{\bibfnamefont{K.~D.} \bibnamefont{Jordan}},
  \bibinfo{author}{\bibfnamefont{J.}~\bibnamefont{Nichols}}, \bibnamefont{and}
  \bibinfo{author}{\bibfnamefont{P.}~\bibnamefont{Nachtigall}},
  \bibinfo{journal}{Theor. Chem. Acc.} \textbf{\bibinfo{volume}{99}},
  \bibinfo{pages}{135} (\bibinfo{year}{1998}).

\bibitem[{\citenamefont{Grafenstein and Cremer}(2000)}]{GraCre-CPL-00}
\bibinfo{author}{\bibfnamefont{J.}~\bibnamefont{Grafenstein}} \bibnamefont{and}
  \bibinfo{author}{\bibfnamefont{D.}~\bibnamefont{Cremer}},
  \bibinfo{journal}{Chem. Phys. Lett.} \textbf{\bibinfo{volume}{{316}}},
  \bibinfo{pages}{569} (\bibinfo{year}{2000}).

\bibitem[{\citenamefont{Gr\"afenstein and Cremer}(2000)}]{GraCre-PCCP-00}
\bibinfo{author}{\bibfnamefont{J.}~\bibnamefont{Gr\"afenstein}}
  \bibnamefont{and} \bibinfo{author}{\bibfnamefont{D.}~\bibnamefont{Cremer}},
  \bibinfo{journal}{Phys. Chem. Chem. Phys.} \textbf{\bibinfo{volume}{2}},
  \bibinfo{pages}{2091} (\bibinfo{year}{2000}).

\bibitem[{\citenamefont{Gr\"afenstein and Cremer}(2005)}]{GraCre-MP-05}
\bibinfo{author}{\bibfnamefont{J.}~\bibnamefont{Gr\"afenstein}}
  \bibnamefont{and} \bibinfo{author}{\bibfnamefont{D.}~\bibnamefont{Cremer}},
  \bibinfo{journal}{Mol. Phys.} \textbf{\bibinfo{volume}{103}},
  \bibinfo{pages}{279} (\bibinfo{year}{2005}).

\bibitem[{\citenamefont{Takeda et~al.}(2002)\citenamefont{Takeda, Yamanaka, and
  Yamaguchi}}]{TakYamYam-CPL-02}
\bibinfo{author}{\bibfnamefont{R.}~\bibnamefont{Takeda}},
  \bibinfo{author}{\bibfnamefont{S.}~\bibnamefont{Yamanaka}}, \bibnamefont{and}
  \bibinfo{author}{\bibfnamefont{K.}~\bibnamefont{Yamaguchi}},
  \bibinfo{journal}{Chem. Phys. Lett.} \textbf{\bibinfo{volume}{{366}}},
  \bibinfo{pages}{321} (\bibinfo{year}{2002}).

\bibitem[{\citenamefont{Nakata et~al.}(2006)\citenamefont{Nakata, Ukai,
  Yamanaka, Takada, and Yamaguchi}}]{NakUkaYamTakYam-IJQC-06}
\bibinfo{author}{\bibfnamefont{K.}~\bibnamefont{Nakata}},
  \bibinfo{author}{\bibfnamefont{T.}~\bibnamefont{Ukai}},
  \bibinfo{author}{\bibfnamefont{S.}~\bibnamefont{Yamanaka}},
  \bibinfo{author}{\bibfnamefont{T.}~\bibnamefont{Takada}}, \bibnamefont{and}
  \bibinfo{author}{\bibfnamefont{K.}~\bibnamefont{Yamaguchi}},
  \bibinfo{journal}{Int. J. Quantum. Chem.} \textbf{\bibinfo{volume}{{106}}},
  \bibinfo{pages}{3325} (\bibinfo{year}{2006}).

\bibitem[{\citenamefont{Yamanaka et~al.}(2006)\citenamefont{Yamanaka, Nakata,
  Ukai, Takada, and Yamaguchi}}]{YamNakUkaTakYam-IJQC-06}
\bibinfo{author}{\bibfnamefont{S.}~\bibnamefont{Yamanaka}},
  \bibinfo{author}{\bibfnamefont{K.}~\bibnamefont{Nakata}},
  \bibinfo{author}{\bibfnamefont{T.}~\bibnamefont{Ukai}},
  \bibinfo{author}{\bibfnamefont{T.}~\bibnamefont{Takada}}, \bibnamefont{and}
  \bibinfo{author}{\bibfnamefont{K.}~\bibnamefont{Yamaguchi}},
  \bibinfo{journal}{Int. J. Quantum. Chem.} \textbf{\bibinfo{volume}{{106}}},
  \bibinfo{pages}{3312} (\bibinfo{year}{2006}).

\bibitem[{\citenamefont{Ukai et~al.}(2007)\citenamefont{Ukai, Nakata, Yamanaka,
  Takada, and Yamaguchi}}]{UkaNakYamTakYam-MP-07}
\bibinfo{author}{\bibfnamefont{T.}~\bibnamefont{Ukai}},
  \bibinfo{author}{\bibfnamefont{K.}~\bibnamefont{Nakata}},
  \bibinfo{author}{\bibfnamefont{S.}~\bibnamefont{Yamanaka}},
  \bibinfo{author}{\bibfnamefont{T.}~\bibnamefont{Takada}}, \bibnamefont{and}
  \bibinfo{author}{\bibfnamefont{K.}~\bibnamefont{Yamaguchi}},
  \bibinfo{journal}{Mol. Phys.} \textbf{\bibinfo{volume}{{105}}},
  \bibinfo{pages}{2667} (\bibinfo{year}{2007}).

\bibitem[{\citenamefont{Gusarov et~al.}(2004)\citenamefont{Gusarov, Malmqvist,
  Lindh, and Roos}}]{GusMalLinRoo-TCA-04}
\bibinfo{author}{\bibfnamefont{S.}~\bibnamefont{Gusarov}},
  \bibinfo{author}{\bibfnamefont{P.-A.} \bibnamefont{Malmqvist}},
  \bibinfo{author}{\bibfnamefont{R.}~\bibnamefont{Lindh}}, \bibnamefont{and}
  \bibinfo{author}{\bibfnamefont{B.~O.} \bibnamefont{Roos}},
  \bibinfo{journal}{Theor. Chim. Acc.} \textbf{\bibinfo{volume}{112}},
  \bibinfo{pages}{84} (\bibinfo{year}{2004}).

\bibitem[{\citenamefont{P\'erez-Jim\'enez and
  P\'erez-Jord\'a}(2007)}]{PerPer-PRA-07}
\bibinfo{author}{\bibfnamefont{A.~J.} \bibnamefont{P\'erez-Jim\'enez}}
  \bibnamefont{and} \bibinfo{author}{\bibfnamefont{J.~M.}
  \bibnamefont{P\'erez-Jord\'a}}, \bibinfo{journal}{Phys. Rev. A}
  \textbf{\bibinfo{volume}{75}}, \bibinfo{pages}{012503}
  (\bibinfo{year}{2007}).

\bibitem[{\citenamefont{P\'erez-Jim\'enez
  et~al.}(2007{\natexlab{a}})\citenamefont{P\'erez-Jim\'enez, P\'erez-Jord\'a,
  Moreira, and Illas}}]{PerPerMorIll-JCC-07}
\bibinfo{author}{\bibfnamefont{A.~J.} \bibnamefont{P\'erez-Jim\'enez}},
  \bibinfo{author}{\bibfnamefont{J.~M.} \bibnamefont{P\'erez-Jord\'a}},
  \bibinfo{author}{\bibfnamefont{I.~D. P.~R.} \bibnamefont{Moreira}},
  \bibnamefont{and} \bibinfo{author}{\bibfnamefont{F.}~\bibnamefont{Illas}},
  \bibinfo{journal}{J. Comput. Chem.} \textbf{\bibinfo{volume}{28}},
  \bibinfo{pages}{2559} (\bibinfo{year}{2007}{\natexlab{a}}).

\bibitem[{\citenamefont{P\'erez-Jim\'enez
  et~al.}(2007{\natexlab{b}})\citenamefont{P\'erez-Jim\'enez, P\'erez-Jord\'a,
  and Sancho-Garc\'ia}}]{PerPerSan-JCP-07}
\bibinfo{author}{\bibfnamefont{A.~J.} \bibnamefont{P\'erez-Jim\'enez}},
  \bibinfo{author}{\bibfnamefont{J.~M.} \bibnamefont{P\'erez-Jord\'a}},
  \bibnamefont{and} \bibinfo{author}{\bibfnamefont{J.~C.}
  \bibnamefont{Sancho-Garc\'ia}}, \bibinfo{journal}{J. Chem. Phys.}
  \textbf{\bibinfo{volume}{127}}, \bibinfo{pages}{104102}
  (\bibinfo{year}{2007}{\natexlab{b}}).

\bibitem[{\citenamefont{Weimer et~al.}(2008)\citenamefont{Weimer, Sala, and
  G\"orling}}]{WeiDelGor-JCP-08}
\bibinfo{author}{\bibfnamefont{M.}~\bibnamefont{Weimer}},
  \bibinfo{author}{\bibfnamefont{F.~D.} \bibnamefont{Sala}}, \bibnamefont{and}
  \bibinfo{author}{\bibfnamefont{A.}~\bibnamefont{G\"orling}},
  \bibinfo{journal}{J. Chem. Phys.} \textbf{\bibinfo{volume}{128}},
  \bibinfo{pages}{144109} (\bibinfo{year}{2008}).

\bibitem[{\citenamefont{Kurzweil et~al.}(2009)\citenamefont{Kurzweil, Lawler,
  and Head-Gordon}}]{KurLawHea-MP-09}
\bibinfo{author}{\bibfnamefont{Y.}~\bibnamefont{Kurzweil}},
  \bibinfo{author}{\bibfnamefont{K.~V.} \bibnamefont{Lawler}},
  \bibnamefont{and}
  \bibinfo{author}{\bibfnamefont{M.}~\bibnamefont{Head-Gordon}},
  \bibinfo{journal}{Mol. Phys.} \textbf{\bibinfo{volume}{107}},
  \bibinfo{pages}{2103} (\bibinfo{year}{2009}).

\bibitem[{\citenamefont{Savin and Flad}(1995)}]{SavFla-IJQC-95}
\bibinfo{author}{\bibfnamefont{A.}~\bibnamefont{Savin}} \bibnamefont{and}
  \bibinfo{author}{\bibfnamefont{H.-J.} \bibnamefont{Flad}},
  \bibinfo{journal}{Int. J. Quantum. Chem.} \textbf{\bibinfo{volume}{{56}}},
  \bibinfo{pages}{327} (\bibinfo{year}{1995}).

\bibitem[{\citenamefont{Savin}(1996{\natexlab{a}})}]{Sav-INC-96a}
\bibinfo{author}{\bibfnamefont{A.}~\bibnamefont{Savin}}, in
  \emph{\bibinfo{booktitle}{Recent Advances in Density Functional Theory}},
  edited by \bibinfo{editor}{\bibfnamefont{D.~P.} \bibnamefont{Chong}}
  (\bibinfo{publisher}{World Scientific}, \bibinfo{year}{1996}{\natexlab{a}}).

\bibitem[{\citenamefont{Savin}(1996{\natexlab{b}})}]{Sav-INC-96}
\bibinfo{author}{\bibfnamefont{A.}~\bibnamefont{Savin}}, in
  \emph{\bibinfo{booktitle}{Recent Developments of Modern Density Functional
  Theory}}, edited by \bibinfo{editor}{\bibfnamefont{J.~M.}
  \bibnamefont{Seminario}} (\bibinfo{publisher}{Elsevier},
  \bibinfo{address}{Amsterdam}, \bibinfo{year}{1996}{\natexlab{b}}), pp.
  \bibinfo{pages}{327--357}.

\bibitem[{\citenamefont{Leininger et~al.}(1997)\citenamefont{Leininger, Stoll,
  Werner, and Savin}}]{LeiStoWerSav-CPL-97}
\bibinfo{author}{\bibfnamefont{T.}~\bibnamefont{Leininger}},
  \bibinfo{author}{\bibfnamefont{H.}~\bibnamefont{Stoll}},
  \bibinfo{author}{\bibfnamefont{H.-J.} \bibnamefont{Werner}},
  \bibnamefont{and} \bibinfo{author}{\bibfnamefont{A.}~\bibnamefont{Savin}},
  \bibinfo{journal}{Chem. Phys. Lett.} \textbf{\bibinfo{volume}{{275}}},
  \bibinfo{pages}{151} (\bibinfo{year}{1997}).

\bibitem[{\citenamefont{Pollet et~al.}(2002)\citenamefont{Pollet, Savin,
  Leininger, and Stoll}}]{PolSavLeiSto-JCP-02}
\bibinfo{author}{\bibfnamefont{R.}~\bibnamefont{Pollet}},
  \bibinfo{author}{\bibfnamefont{A.}~\bibnamefont{Savin}},
  \bibinfo{author}{\bibfnamefont{T.}~\bibnamefont{Leininger}},
  \bibnamefont{and} \bibinfo{author}{\bibfnamefont{H.}~\bibnamefont{Stoll}},
  \bibinfo{journal}{J. Chem. Phys.} \textbf{\bibinfo{volume}{{116}}},
  \bibinfo{pages}{1250} (\bibinfo{year}{2002}).

\bibitem[{\citenamefont{Savin et~al.}(2003)\citenamefont{Savin, Colonna, and
  Pollet}}]{SavColPol-IJQC-03}
\bibinfo{author}{\bibfnamefont{A.}~\bibnamefont{Savin}},
  \bibinfo{author}{\bibfnamefont{F.}~\bibnamefont{Colonna}}, \bibnamefont{and}
  \bibinfo{author}{\bibfnamefont{R.}~\bibnamefont{Pollet}},
  \bibinfo{journal}{Int. J. Quantum. Chem.} \textbf{\bibinfo{volume}{{93}}},
  \bibinfo{pages}{166} (\bibinfo{year}{2003}).

\bibitem[{\citenamefont{Toulouse et~al.}(2004)\citenamefont{Toulouse, Colonna,
  and Savin}}]{TouColSav-PRA-04}
\bibinfo{author}{\bibfnamefont{J.}~\bibnamefont{Toulouse}},
  \bibinfo{author}{\bibfnamefont{F.}~\bibnamefont{Colonna}}, \bibnamefont{and}
  \bibinfo{author}{\bibfnamefont{A.}~\bibnamefont{Savin}},
  \bibinfo{journal}{Phys. Rev. A} \textbf{\bibinfo{volume}{70}},
  \bibinfo{pages}{062505} (\bibinfo{year}{2004}).

\bibitem[{\citenamefont{Pedersen and Jensen}()}]{PedJen-JJJ-XX}
\bibinfo{author}{\bibfnamefont{J.~K.} \bibnamefont{Pedersen}} \bibnamefont{and}
  \bibinfo{author}{\bibfnamefont{H.~J.~A.} \bibnamefont{Jensen}},
  \bibinfo{note}{a second order MCSCF-DFT hybrid algorithm (unpublished)}.

\bibitem[{\citenamefont{Fromager et~al.}(2007)\citenamefont{Fromager, Toulouse,
  and Jensen}}]{FroTouJen-JCP-07}
\bibinfo{author}{\bibfnamefont{E.}~\bibnamefont{Fromager}},
  \bibinfo{author}{\bibfnamefont{J.}~\bibnamefont{Toulouse}}, \bibnamefont{and}
  \bibinfo{author}{\bibfnamefont{H.~J.~A.} \bibnamefont{Jensen}},
  \bibinfo{journal}{J. Chem. Phys.} \textbf{\bibinfo{volume}{126}},
  \bibinfo{pages}{074111} (\bibinfo{year}{2007}).

\bibitem[{\citenamefont{Fromager et~al.}(2009)\citenamefont{Fromager, R\'eal,
  W{\aa}hlin, Wahlgren, and Jensen}}]{FroReaWahWahJen-JCP-09}
\bibinfo{author}{\bibfnamefont{E.}~\bibnamefont{Fromager}},
  \bibinfo{author}{\bibfnamefont{F.}~\bibnamefont{R\'eal}},
  \bibinfo{author}{\bibfnamefont{P.}~\bibnamefont{W{\aa}hlin}},
  \bibinfo{author}{\bibfnamefont{U.}~\bibnamefont{Wahlgren}}, \bibnamefont{and}
  \bibinfo{author}{\bibfnamefont{H.~J.~A.} \bibnamefont{Jensen}},
  \bibinfo{journal}{J. Chem. Phys.} \textbf{\bibinfo{volume}{131}},
  \bibinfo{pages}{054107} (\bibinfo{year}{2009}).

\bibitem[{\citenamefont{Fromager et~al.}(2010)\citenamefont{Fromager,
  Cimiraglia, and Jensen}}]{FroCimJen-PRA-10}
\bibinfo{author}{\bibfnamefont{E.}~\bibnamefont{Fromager}},
  \bibinfo{author}{\bibfnamefont{R.}~\bibnamefont{Cimiraglia}},
  \bibnamefont{and} \bibinfo{author}{\bibfnamefont{H.~J.~A.}
  \bibnamefont{Jensen}}, \bibinfo{journal}{Phys. Rev. A}
  \textbf{\bibinfo{volume}{81}}, \bibinfo{pages}{024502}
  (\bibinfo{year}{2010}).

\bibitem[{\citenamefont{Tsuchimochi et~al.}(2010)\citenamefont{Tsuchimochi,
  Scuseria, and Savin}}]{TsuScuSav-JCP-10}
\bibinfo{author}{\bibfnamefont{T.}~\bibnamefont{Tsuchimochi}},
  \bibinfo{author}{\bibfnamefont{G.~E.} \bibnamefont{Scuseria}},
  \bibnamefont{and} \bibinfo{author}{\bibfnamefont{A.}~\bibnamefont{Savin}},
  \bibinfo{journal}{J. Chem. Phys.} \textbf{\bibinfo{volume}{132}},
  \bibinfo{pages}{024111} (\bibinfo{year}{2010}).

\bibitem[{\citenamefont{Tsuchimochi and Scuseria}(2011)}]{TsuScu-JCP-11}
\bibinfo{author}{\bibfnamefont{T.}~\bibnamefont{Tsuchimochi}} \bibnamefont{and}
  \bibinfo{author}{\bibfnamefont{G.~E.} \bibnamefont{Scuseria}},
  \bibinfo{journal}{J. Chem. Phys.} \textbf{\bibinfo{volume}{134}},
  \bibinfo{pages}{064101} (\bibinfo{year}{2011}).

\bibitem[{\citenamefont{Pernal}(2010)}]{Per-PRA-10}
\bibinfo{author}{\bibfnamefont{K.}~\bibnamefont{Pernal}},
  \bibinfo{journal}{Phys. Rev. A} \textbf{\bibinfo{volume}{81}},
  \bibinfo{pages}{052511} (\bibinfo{year}{2010}).

\bibitem[{\citenamefont{Rohr et~al.}(2010)\citenamefont{Rohr, Toulouse, and
  Pernal}}]{RohTouPer-PRA-10}
\bibinfo{author}{\bibfnamefont{D.~R.} \bibnamefont{Rohr}},
  \bibinfo{author}{\bibfnamefont{J.}~\bibnamefont{Toulouse}}, \bibnamefont{and}
  \bibinfo{author}{\bibfnamefont{K.}~\bibnamefont{Pernal}},
  \bibinfo{journal}{Phys. Rev. A} \textbf{\bibinfo{volume}{82}},
  \bibinfo{pages}{052502} (\bibinfo{year}{2010}).

\bibitem[{\citenamefont{Rohr and Pernal}(2011)}]{RohPer-JCP-11}
\bibinfo{author}{\bibfnamefont{D.~R.} \bibnamefont{Rohr}} \bibnamefont{and}
  \bibinfo{author}{\bibfnamefont{K.}~\bibnamefont{Pernal}},
  \bibinfo{journal}{J. Chem. Phys.} \textbf{\bibinfo{volume}{135}},
  \bibinfo{pages}{074104} (\bibinfo{year}{2011}).

\bibitem[{\citenamefont{Sharkas et~al.}(2011)\citenamefont{Sharkas, Toulouse,
  and Savin}}]{ShaTouSav-JCP-11}
\bibinfo{author}{\bibfnamefont{K.}~\bibnamefont{Sharkas}},
  \bibinfo{author}{\bibfnamefont{J.}~\bibnamefont{Toulouse}}, \bibnamefont{and}
  \bibinfo{author}{\bibfnamefont{A.}~\bibnamefont{Savin}}, \bibinfo{journal}{J.
  Chem. Phys.} \textbf{\bibinfo{volume}{134}}, \bibinfo{pages}{064113}
  (\bibinfo{year}{2011}).

\bibitem[{\citenamefont{\'Angy\'an et~al.}(2005)\citenamefont{\'Angy\'an,
  Gerber, Savin, and Toulouse}}]{AngGerSavTou-PRA-05}
\bibinfo{author}{\bibfnamefont{J.~G.} \bibnamefont{\'Angy\'an}},
  \bibinfo{author}{\bibfnamefont{I.~C.} \bibnamefont{Gerber}},
  \bibinfo{author}{\bibfnamefont{A.}~\bibnamefont{Savin}}, \bibnamefont{and}
  \bibinfo{author}{\bibfnamefont{J.}~\bibnamefont{Toulouse}},
  \bibinfo{journal}{Phys. Rev. A} \textbf{\bibinfo{volume}{72}},
  \bibinfo{pages}{012510} (\bibinfo{year}{2005}).

\bibitem[{\citenamefont{Levy and Perdew}(1985)}]{LevPer-PRA-85}
\bibinfo{author}{\bibfnamefont{M.}~\bibnamefont{Levy}} \bibnamefont{and}
  \bibinfo{author}{\bibfnamefont{J.~P.} \bibnamefont{Perdew}},
  \bibinfo{journal}{Phys. Rev. A} \textbf{\bibinfo{volume}{32}},
  \bibinfo{pages}{2010} (\bibinfo{year}{1985}).

\bibitem[{\citenamefont{Levy et~al.}(1985)\citenamefont{Levy, Yang, and
  Parr}}]{LevYanPar-JCP-85}
\bibinfo{author}{\bibfnamefont{M.}~\bibnamefont{Levy}},
  \bibinfo{author}{\bibfnamefont{W.}~\bibnamefont{Yang}}, \bibnamefont{and}
  \bibinfo{author}{\bibfnamefont{R.~G.} \bibnamefont{Parr}},
  \bibinfo{journal}{J. Chem. Phys.} \textbf{\bibinfo{volume}{83}},
  \bibinfo{pages}{2334} (\bibinfo{year}{1985}).

\bibitem[{\citenamefont{Levy}(1991)}]{Lev-PRA-91}
\bibinfo{author}{\bibfnamefont{M.}~\bibnamefont{Levy}}, \bibinfo{journal}{Phys.
  Rev. A} \textbf{\bibinfo{volume}{43}}, \bibinfo{pages}{4637}
  (\bibinfo{year}{1991}).

\bibitem[{\citenamefont{Levy and Perdew}(1993)}]{LevPer-PRB-93}
\bibinfo{author}{\bibfnamefont{M.}~\bibnamefont{Levy}} \bibnamefont{and}
  \bibinfo{author}{\bibfnamefont{J.~P.} \bibnamefont{Perdew}},
  \bibinfo{journal}{Phys. Rev. B} \textbf{\bibinfo{volume}{48}},
  \bibinfo{pages}{11638} (\bibinfo{year}{1993}).

\bibitem[{\citenamefont{Becke}(1996)}]{Bec-JCP-96}
\bibinfo{author}{\bibfnamefont{A.~D.} \bibnamefont{Becke}},
  \bibinfo{journal}{J. Chem. Phys.} \textbf{\bibinfo{volume}{104}},
  \bibinfo{pages}{1040} (\bibinfo{year}{1996}).

\bibitem[{\citenamefont{Ernzerhof et~al.}(1996)\citenamefont{Ernzerhof, Perdew,
  and Burke}}]{ErnPerBur-INC-96}
\bibinfo{author}{\bibfnamefont{M.}~\bibnamefont{Ernzerhof}},
  \bibinfo{author}{\bibfnamefont{J.~P.} \bibnamefont{Perdew}},
  \bibnamefont{and} \bibinfo{author}{\bibfnamefont{K.}~\bibnamefont{Burke}}, in
  \emph{\bibinfo{booktitle}{Density Functional Theory}}, edited by
  \bibinfo{editor}{\bibfnamefont{R.}~\bibnamefont{Nalewajski}}
  (\bibinfo{publisher}{Springer-Verlag}, \bibinfo{address}{Berlin},
  \bibinfo{year}{1996}).

\bibitem[{\citenamefont{Fromager and Jensen}(2008)}]{FroJen-PRA-08}
\bibinfo{author}{\bibfnamefont{E.}~\bibnamefont{Fromager}} \bibnamefont{and}
  \bibinfo{author}{\bibfnamefont{H.~J.~A.} \bibnamefont{Jensen}},
  \bibinfo{journal}{Phys. Rev. A} \textbf{\bibinfo{volume}{78}},
  \bibinfo{pages}{022504} (\bibinfo{year}{2008}).

\bibitem[{\citenamefont{\'Angy\'an}(2008)}]{Ang-PRA-08}
\bibinfo{author}{\bibfnamefont{J.~G.} \bibnamefont{\'Angy\'an}},
  \bibinfo{journal}{Phys. Rev. A} \textbf{\bibinfo{volume}{78}},
  \bibinfo{pages}{022510} (\bibinfo{year}{2008}).

\bibitem[{\citenamefont{Grimme}(2006)}]{Gri-JCP-06}
\bibinfo{author}{\bibfnamefont{S.}~\bibnamefont{Grimme}}, \bibinfo{journal}{J.
  Chem. Phys.} \textbf{\bibinfo{volume}{124}}, \bibinfo{pages}{034108}
  (\bibinfo{year}{2006}).

\bibitem[{\citenamefont{Zhao et~al.}(2009)\citenamefont{Zhao, Tishchenko, Gour,
  Li, Lutz, Piecuch, and Truhlar}}]{ZhaTisGouLiLutPieTru-JPCA-09}
\bibinfo{author}{\bibfnamefont{Y.}~\bibnamefont{Zhao}},
  \bibinfo{author}{\bibfnamefont{O.}~\bibnamefont{Tishchenko}},
  \bibinfo{author}{\bibfnamefont{J.~R.} \bibnamefont{Gour}},
  \bibinfo{author}{\bibfnamefont{W.}~\bibnamefont{Li}},
  \bibinfo{author}{\bibfnamefont{J.~J.} \bibnamefont{Lutz}},
  \bibinfo{author}{\bibfnamefont{P.}~\bibnamefont{Piecuch}}, \bibnamefont{and}
  \bibinfo{author}{\bibfnamefont{D.~G.} \bibnamefont{Truhlar}},
  \bibinfo{journal}{J. Phys. Chem. A} \textbf{\bibinfo{volume}{113}},
  \bibinfo{pages}{5786} (\bibinfo{year}{2009}).

\bibitem[{\citenamefont{Frisch et~al.}()\citenamefont{Frisch, Trucks, Schlegel,
  Scuseria, Robb, Cheeseman, Scalmani, Barone, Mennucci, Petersson
  et~al.}}]{Gaussian-PROG-09}
\bibinfo{author}{\bibfnamefont{M.~J.} \bibnamefont{Frisch}},
  \bibinfo{author}{\bibfnamefont{G.~W.} \bibnamefont{Trucks}},
  \bibinfo{author}{\bibfnamefont{H.~B.} \bibnamefont{Schlegel}},
  \bibinfo{author}{\bibfnamefont{G.~E.} \bibnamefont{Scuseria}},
  \bibinfo{author}{\bibfnamefont{M.~A.} \bibnamefont{Robb}},
  \bibinfo{author}{\bibfnamefont{J.~R.} \bibnamefont{Cheeseman}},
  \bibinfo{author}{\bibfnamefont{G.}~\bibnamefont{Scalmani}},
  \bibinfo{author}{\bibfnamefont{V.}~\bibnamefont{Barone}},
  \bibinfo{author}{\bibfnamefont{B.}~\bibnamefont{Mennucci}},
  \bibinfo{author}{\bibfnamefont{G.~A.} \bibnamefont{Petersson}},
  \bibnamefont{et~al.}, \emph{\bibinfo{title}{Gaussian~09 {R}evision {A}.1}},
  \bibinfo{note}{\uppercase{G}aussian Inc. Wallingford CT 2009}.

\bibitem[{\citenamefont{Angeli et~al.}()\citenamefont{Angeli, Bak, Bakken,
  Christiansen, Cimiraglia, Coriani, Dahle, Dalskov, Enevoldsen, Fernandez
  et~al.}}]{Dal-PROG-11}
\bibinfo{author}{\bibfnamefont{C.}~\bibnamefont{Angeli}},
  \bibinfo{author}{\bibfnamefont{K.~L.} \bibnamefont{Bak}},
  \bibinfo{author}{\bibfnamefont{V.}~\bibnamefont{Bakken}},
  \bibinfo{author}{\bibfnamefont{O.}~\bibnamefont{Christiansen}},
  \bibinfo{author}{\bibfnamefont{R.}~\bibnamefont{Cimiraglia}},
  \bibinfo{author}{\bibfnamefont{S.}~\bibnamefont{Coriani}},
  \bibinfo{author}{\bibfnamefont{P.}~\bibnamefont{Dahle}},
  \bibinfo{author}{\bibfnamefont{E.~K.} \bibnamefont{Dalskov}},
  \bibinfo{author}{\bibfnamefont{T.}~\bibnamefont{Enevoldsen}},
  \bibinfo{author}{\bibfnamefont{B.}~\bibnamefont{Fernandez}},
  \bibnamefont{et~al.}, \emph{\bibinfo{title}{Dalton, a molecular electronic
  structure program, release dalton2011 (2011), see http://daltonprogram.org}}.

\bibitem[{\citenamefont{Pedersen}(2004)}]{Ped-THESIS-04}
\bibinfo{author}{\bibfnamefont{J.~K.} \bibnamefont{Pedersen}},
  \bibinfo{type}{{PhD thesis}}, \bibinfo{school}{University of Southern
  Denmark}, \bibinfo{address}{Odense} (\bibinfo{year}{2004}).

\bibitem[{\citenamefont{Jensen and J{\o}rgensen}(1984)}]{JenJor-JCP-84}
\bibinfo{author}{\bibfnamefont{H.~J.~A.} \bibnamefont{Jensen}}
  \bibnamefont{and}
  \bibinfo{author}{\bibfnamefont{P.}~\bibnamefont{J{\o}rgensen}},
  \bibinfo{journal}{J. Chem. Phys.} \textbf{\bibinfo{volume}{{80}}},
  \bibinfo{pages}{1204} (\bibinfo{year}{1984}).

\bibitem[{\citenamefont{Jensen and Agren}(1984)}]{JenAgr-CPL-84}
\bibinfo{author}{\bibfnamefont{H.~J.~A.} \bibnamefont{Jensen}}
  \bibnamefont{and} \bibinfo{author}{\bibfnamefont{H.}~\bibnamefont{Agren}},
  \bibinfo{journal}{Chem. Phys. Lett.} \textbf{\bibinfo{volume}{110}},
  \bibinfo{pages}{140} (\bibinfo{year}{1984}).

\bibitem[{\citenamefont{Jensen and Agren}(1986)}]{JenAgr-CP-86}
\bibinfo{author}{\bibfnamefont{H.~J.~A.} \bibnamefont{Jensen}}
  \bibnamefont{and} \bibinfo{author}{\bibfnamefont{H.}~\bibnamefont{Agren}},
  \bibinfo{journal}{Chem. Phys.} \textbf{\bibinfo{volume}{104}},
  \bibinfo{pages}{229} (\bibinfo{year}{1986}).

\bibitem[{\citenamefont{Jensen et~al.}(1987)\citenamefont{Jensen, Jorgensen,
  and Agren}}]{JenJorAgr-JCP-87}
\bibinfo{author}{\bibfnamefont{H.~J.~A.} \bibnamefont{Jensen}},
  \bibinfo{author}{\bibfnamefont{P.}~\bibnamefont{Jorgensen}},
  \bibnamefont{and} \bibinfo{author}{\bibfnamefont{H.}~\bibnamefont{Agren}},
  \bibinfo{journal}{J. Chem. Phys.} \textbf{\bibinfo{volume}{{87}}},
  \bibinfo{pages}{451} (\bibinfo{year}{1987}).

\bibitem[{\citenamefont{Jensen et~al.}(1988)\citenamefont{Jensen, J{\o}rgensen,
  Agren, and Olsen}}]{JenJorAgrOls-JCP-88}
\bibinfo{author}{\bibfnamefont{H.~J.~A.} \bibnamefont{Jensen}},
  \bibinfo{author}{\bibfnamefont{P.}~\bibnamefont{J{\o}rgensen}},
  \bibinfo{author}{\bibfnamefont{H.}~\bibnamefont{Agren}}, \bibnamefont{and}
  \bibinfo{author}{\bibfnamefont{J.}~\bibnamefont{Olsen}}, \bibinfo{journal}{J.
  Chem. Phys.} \textbf{\bibinfo{volume}{{88}}}, \bibinfo{pages}{3834}
  (\bibinfo{year}{1988}).

\bibitem[{\citenamefont{Jensen}(1994)}]{Jen-INC-94}
\bibinfo{author}{\bibfnamefont{H.~J.~A.} \bibnamefont{Jensen}}, in
  \emph{\bibinfo{booktitle}{Relativistic and Electron Correlation Effects in
  Molecules and Solids}}, edited by \bibinfo{editor}{\bibfnamefont{G.~L.}
  \bibnamefont{Malli}} (\bibinfo{publisher}{Plenum}, \bibinfo{address}{New
  York}, \bibinfo{year}{1994}), p. \bibinfo{pages}{179}.

\bibitem[{\citenamefont{Perdew et~al.}(1996{\natexlab{a}})\citenamefont{Perdew,
  Burke, and Ernzerhof}}]{PerBurErn-PRL-96}
\bibinfo{author}{\bibfnamefont{J.~P.} \bibnamefont{Perdew}},
  \bibinfo{author}{\bibfnamefont{K.}~\bibnamefont{Burke}}, \bibnamefont{and}
  \bibinfo{author}{\bibfnamefont{M.}~\bibnamefont{Ernzerhof}},
  \bibinfo{journal}{Phys. Rev. Lett.} \textbf{\bibinfo{volume}{77}},
  \bibinfo{pages}{3865} (\bibinfo{year}{1996}{\natexlab{a}}).

\bibitem[{\citenamefont{Becke}(1988)}]{Bec-PRA-88}
\bibinfo{author}{\bibfnamefont{A.~D.} \bibnamefont{Becke}},
  \bibinfo{journal}{Phys. Rev. A} \textbf{\bibinfo{volume}{{38}}},
  \bibinfo{pages}{3098} (\bibinfo{year}{1988}).

\bibitem[{\citenamefont{Lee et~al.}(1988)\citenamefont{Lee, Yang, and
  Parr}}]{LeeYanPar-PRB-88}
\bibinfo{author}{\bibfnamefont{C.}~\bibnamefont{Lee}},
  \bibinfo{author}{\bibfnamefont{W.}~\bibnamefont{Yang}}, \bibnamefont{and}
  \bibinfo{author}{\bibfnamefont{R.~G.} \bibnamefont{Parr}},
  \bibinfo{journal}{Phys. Rev. B} \textbf{\bibinfo{volume}{37}},
  \bibinfo{pages}{785} (\bibinfo{year}{1988}).

\bibitem[{\citenamefont{Goerigk and Grimme}(2010)}]{GoeGri-JCTC-10}
\bibinfo{author}{\bibfnamefont{L.}~\bibnamefont{Goerigk}} \bibnamefont{and}
  \bibinfo{author}{\bibfnamefont{S.}~\bibnamefont{Grimme}},
  \bibinfo{journal}{J. Chem. Theory Comput} \textbf{\bibinfo{volume}{6}},
  \bibinfo{pages}{107} (\bibinfo{year}{2010}).

\bibitem[{\citenamefont{Goerigk and Grimme}(2011)}]{GoeGri-JCTC-11}
\bibinfo{author}{\bibfnamefont{L.}~\bibnamefont{Goerigk}} \bibnamefont{and}
  \bibinfo{author}{\bibfnamefont{S.}~\bibnamefont{Grimme}},
  \bibinfo{journal}{J. Chem. Theory Comput} \textbf{\bibinfo{volume}{7}},
  \bibinfo{pages}{291} (\bibinfo{year}{2011}).

\bibitem[{\citenamefont{Wheeler et~al.}(2007)\citenamefont{Wheeler, Ess, and
  Houk}}]{WheEssHou-JPCA-07}
\bibinfo{author}{\bibfnamefont{S.~E.} \bibnamefont{Wheeler}},
  \bibinfo{author}{\bibfnamefont{D.~H.} \bibnamefont{Ess}}, \bibnamefont{and}
  \bibinfo{author}{\bibfnamefont{K.~N.} \bibnamefont{Houk}},
  \bibinfo{journal}{J. Phys. Chem. A} \textbf{\bibinfo{volume}{112}},
  \bibinfo{pages}{1798} (\bibinfo{year}{2007}).

\bibitem[{\citenamefont{Saito et~al.}(2010)\citenamefont{Saito, Nishihara,
  Kataoka, Nakanishi, Kitagawa, Kawakami, S.Yamanaka, Okumura, and
  Yamaguchi}}]{SaiNisKatNakKitKawYamOkuYam-JPCA-10}
\bibinfo{author}{\bibfnamefont{T.}~\bibnamefont{Saito}},
  \bibinfo{author}{\bibfnamefont{S.}~\bibnamefont{Nishihara}},
  \bibinfo{author}{\bibfnamefont{Y.}~\bibnamefont{Kataoka}},
  \bibinfo{author}{\bibfnamefont{Y.}~\bibnamefont{Nakanishi}},
  \bibinfo{author}{\bibfnamefont{Y.}~\bibnamefont{Kitagawa}},
  \bibinfo{author}{\bibfnamefont{T.}~\bibnamefont{Kawakami}},
  \bibinfo{author}{\bibnamefont{S.Yamanaka}},
  \bibinfo{author}{\bibfnamefont{M.}~\bibnamefont{Okumura}}, \bibnamefont{and}
  \bibinfo{author}{\bibfnamefont{K.}~\bibnamefont{Yamaguchi}},
  \bibinfo{journal}{J. Phys. Chem. A} \textbf{\bibinfo{volume}{114}},
  \bibinfo{pages}{12116} (\bibinfo{year}{2010}).

\bibitem[{\citenamefont{Dunning}(1989)}]{Dun-JCP-89}
\bibinfo{author}{\bibfnamefont{T.~H.} \bibnamefont{Dunning}},
  \bibinfo{journal}{J. Chem. Phys.} \textbf{\bibinfo{volume}{90}},
  \bibinfo{pages}{1007} (\bibinfo{year}{1989}).

\bibitem[{\citenamefont{Kendall et~al.}(1992)\citenamefont{Kendall, Dunning,
  and Harrison}}]{KenDunHar-JCP-92}
\bibinfo{author}{\bibfnamefont{R.~A.} \bibnamefont{Kendall}},
  \bibinfo{author}{\bibfnamefont{T.~H.} \bibnamefont{Dunning}},
  \bibnamefont{and} \bibinfo{author}{\bibfnamefont{R.~J.}
  \bibnamefont{Harrison}}, \bibinfo{journal}{J. Chem. Phys.}
  \textbf{\bibinfo{volume}{96}}, \bibinfo{pages}{6796} (\bibinfo{year}{1992}).

\bibitem[{\citenamefont{Zhao et~al.}(2005)\citenamefont{Zhao, Schultz, and
  Truhlar}}]{ZhaSchTru-JCP-05}
\bibinfo{author}{\bibfnamefont{Y.}~\bibnamefont{Zhao}},
  \bibinfo{author}{\bibfnamefont{N.~E.} \bibnamefont{Schultz}},
  \bibnamefont{and} \bibinfo{author}{\bibfnamefont{D.~G.}
  \bibnamefont{Truhlar}}, \bibinfo{journal}{J. Chem. Phys.}
  \textbf{\bibinfo{volume}{123}}, \bibinfo{pages}{161103}
  (\bibinfo{year}{2005}).

\bibitem[{\citenamefont{Krishnan et~al.}(1980)\citenamefont{Krishnan, Binkley,
  Seeger, and Pople}}]{KriBinSeePop-JCP-80}
\bibinfo{author}{\bibfnamefont{R.}~\bibnamefont{Krishnan}},
  \bibinfo{author}{\bibfnamefont{J.~S.} \bibnamefont{Binkley}},
  \bibinfo{author}{\bibfnamefont{R.}~\bibnamefont{Seeger}}, \bibnamefont{and}
  \bibinfo{author}{\bibfnamefont{J.~A.} \bibnamefont{Pople}},
  \bibinfo{journal}{J. Chem. Phys.} \textbf{\bibinfo{volume}{72}},
  \bibinfo{pages}{650} (\bibinfo{year}{1980}).

\bibitem[{\citenamefont{McLean and Chandler}(1980)}]{MclCha-JCP-80}
\bibinfo{author}{\bibfnamefont{A.~D.} \bibnamefont{McLean}} \bibnamefont{and}
  \bibinfo{author}{\bibfnamefont{G.~S.} \bibnamefont{Chandler}},
  \bibinfo{journal}{J. Chem. Phys.} \textbf{\bibinfo{volume}{72}},
  \bibinfo{pages}{5639} (\bibinfo{year}{1980}).

\bibitem[{\citenamefont{M{\o}ller and Plesset}(1934)}]{MolPle-PR-34}
\bibinfo{author}{\bibfnamefont{C.}~\bibnamefont{M{\o}ller}} \bibnamefont{and}
  \bibinfo{author}{\bibfnamefont{M.~S.} \bibnamefont{Plesset}},
  \bibinfo{journal}{Phys. Rev.} \textbf{\bibinfo{volume}{46}},
  \bibinfo{pages}{618} (\bibinfo{year}{1934}).

\bibitem[{\citenamefont{Hirao}(1992)}]{Hir-CPL-92}
\bibinfo{author}{\bibfnamefont{K.}~\bibnamefont{Hirao}},
  \bibinfo{journal}{Chem. Phys. Lett.} \textbf{\bibinfo{volume}{190}},
  \bibinfo{pages}{374} (\bibinfo{year}{1992}).

\bibitem[{\citenamefont{Adamo and Barone}(1999)}]{AdaBar-JCP-99}
\bibinfo{author}{\bibfnamefont{C.}~\bibnamefont{Adamo}} \bibnamefont{and}
  \bibinfo{author}{\bibfnamefont{V.}~\bibnamefont{Barone}},
  \bibinfo{journal}{J. Chem. Phys.} \textbf{\bibinfo{volume}{110}},
  \bibinfo{pages}{6158} (\bibinfo{year}{1999}).

\bibitem[{\citenamefont{Adamo and Barone}(1998)}]{AdaBar-CPL-98}
\bibinfo{author}{\bibfnamefont{C.}~\bibnamefont{Adamo}} \bibnamefont{and}
  \bibinfo{author}{\bibfnamefont{V.}~\bibnamefont{Barone}},
  \bibinfo{journal}{Chem. Phys. Lett.} \textbf{\bibinfo{volume}{298}},
  \bibinfo{pages}{113} (\bibinfo{year}{1998}).

\bibitem[{\citenamefont{Adamo and Barone}(1997)}]{AdaBar-CPL-97}
\bibinfo{author}{\bibfnamefont{C.}~\bibnamefont{Adamo}} \bibnamefont{and}
  \bibinfo{author}{\bibfnamefont{V.}~\bibnamefont{Barone}},
  \bibinfo{journal}{Chem. Phys. Lett.} \textbf{\bibinfo{volume}{274}},
  \bibinfo{pages}{242} (\bibinfo{year}{1997}).

\bibitem[{\citenamefont{Becke}(1993)}]{Bec-JCP-93}
\bibinfo{author}{\bibfnamefont{A.~D.} \bibnamefont{Becke}},
  \bibinfo{journal}{J. Chem. Phys.} \textbf{\bibinfo{volume}{98}},
  \bibinfo{pages}{5648} (\bibinfo{year}{1993}).

\bibitem[{\citenamefont{Barone and Adamo}(1994)}]{BarAda-CPL-94}
\bibinfo{author}{\bibfnamefont{V.}~\bibnamefont{Barone}} \bibnamefont{and}
  \bibinfo{author}{\bibfnamefont{C.}~\bibnamefont{Adamo}},
  \bibinfo{journal}{Chem. Phys. Lett.} \textbf{\bibinfo{volume}{224}},
  \bibinfo{pages}{432} (\bibinfo{year}{1994}).

\bibitem[{\citenamefont{Goll et~al.}(2006)\citenamefont{Goll, Werner, Stoll,
  Leininger, Gori-Giorgi, and Savin}}]{GolWerStoLeiGorSav-CP-06}
\bibinfo{author}{\bibfnamefont{E.}~\bibnamefont{Goll}},
  \bibinfo{author}{\bibfnamefont{H.-J.} \bibnamefont{Werner}},
  \bibinfo{author}{\bibfnamefont{H.}~\bibnamefont{Stoll}},
  \bibinfo{author}{\bibfnamefont{T.}~\bibnamefont{Leininger}},
  \bibinfo{author}{\bibfnamefont{P.}~\bibnamefont{Gori-Giorgi}},
  \bibnamefont{and} \bibinfo{author}{\bibfnamefont{A.}~\bibnamefont{Savin}},
  \bibinfo{journal}{Chem. Phys.} \textbf{\bibinfo{volume}{329}},
  \bibinfo{pages}{276} (\bibinfo{year}{2006}).

\bibitem[{\citenamefont{Perdew et~al.}(1996{\natexlab{b}})\citenamefont{Perdew,
  Ernzerhof, and Burke}}]{PerErnBur-JCP-96}
\bibinfo{author}{\bibfnamefont{J.~P.} \bibnamefont{Perdew}},
  \bibinfo{author}{\bibfnamefont{M.}~\bibnamefont{Ernzerhof}},
  \bibnamefont{and} \bibinfo{author}{\bibfnamefont{K.}~\bibnamefont{Burke}},
  \bibinfo{journal}{J. Chem. Phys} \textbf{\bibinfo{volume}{105}},
  \bibinfo{pages}{9982} (\bibinfo{year}{1996}{\natexlab{b}}).

\bibitem[{\citenamefont{Perdew et~al.}(1995)\citenamefont{Perdew, Savin, and
  Burke}}]{PerSavBur-PRA-95}
\bibinfo{author}{\bibfnamefont{J.~P.} \bibnamefont{Perdew}},
  \bibinfo{author}{\bibfnamefont{A.}~\bibnamefont{Savin}}, \bibnamefont{and}
  \bibinfo{author}{\bibfnamefont{K.}~\bibnamefont{Burke}},
  \bibinfo{journal}{Phys. Rev. A} \textbf{\bibinfo{volume}{51}},
  \bibinfo{pages}{4531} (\bibinfo{year}{1995}).

\bibitem[{\citenamefont{Dewar and Kelemen}(1971)}]{DewKel-JCE-71}
\bibinfo{author}{\bibfnamefont{M.~J.~S.} \bibnamefont{Dewar}} \bibnamefont{and}
  \bibinfo{author}{\bibfnamefont{J.}~\bibnamefont{Kelemen}},
  \bibinfo{journal}{J. Chem. Educ.} \textbf{\bibinfo{volume}{48}},
  \bibinfo{pages}{494} (\bibinfo{year}{1971}).

\bibitem[{\citenamefont{Iikura et~al.}(2001)\citenamefont{Iikura, Tsuneda,
  Yanai, and Hirao}}]{IikTsuYanHir-JCP-01}
\bibinfo{author}{\bibfnamefont{H.}~\bibnamefont{Iikura}},
  \bibinfo{author}{\bibfnamefont{T.}~\bibnamefont{Tsuneda}},
  \bibinfo{author}{\bibfnamefont{T.}~\bibnamefont{Yanai}}, \bibnamefont{and}
  \bibinfo{author}{\bibfnamefont{K.}~\bibnamefont{Hirao}}, \bibinfo{journal}{J.
  Chem. Phys.} \textbf{\bibinfo{volume}{{115}}}, \bibinfo{pages}{3540}
  (\bibinfo{year}{2001}).

\bibitem[{\citenamefont{Yanai et~al.}(2004)\citenamefont{Yanai, Tew, and
  Handy}}]{YanTewHan-CPL-04}
\bibinfo{author}{\bibfnamefont{T.}~\bibnamefont{Yanai}},
  \bibinfo{author}{\bibfnamefont{D.~P.} \bibnamefont{Tew}}, \bibnamefont{and}
  \bibinfo{author}{\bibfnamefont{N.~C.} \bibnamefont{Handy}},
  \bibinfo{journal}{Chem. Phys. Lett.} \textbf{\bibinfo{volume}{393}},
  \bibinfo{pages}{51} (\bibinfo{year}{2004}).

\bibitem[{\citenamefont{Gerber and \'Angy\'an}(2005)}]{GerAng-CPL-05a}
\bibinfo{author}{\bibfnamefont{I.~C.} \bibnamefont{Gerber}} \bibnamefont{and}
  \bibinfo{author}{\bibfnamefont{J.~G.} \bibnamefont{\'Angy\'an}},
  \bibinfo{journal}{Chem. Phys. Lett.} \textbf{\bibinfo{volume}{415}},
  \bibinfo{pages}{100} (\bibinfo{year}{2005}).

\bibitem[{\citenamefont{Vydrov and Scuseria}(2006)}]{VydScu-JCP-06}
\bibinfo{author}{\bibfnamefont{O.~A.} \bibnamefont{Vydrov}} \bibnamefont{and}
  \bibinfo{author}{\bibfnamefont{G.~E.} \bibnamefont{Scuseria}},
  \bibinfo{journal}{J. Chem. Phys.} \textbf{\bibinfo{volume}{125}},
  \bibinfo{pages}{234109} (\bibinfo{year}{2006}).

\bibitem[{\citenamefont{Heyd et~al.}(2003)\citenamefont{Heyd, Scuseria, and
  Ernzerhof}}]{HeyScuErn-JCP-03}
\bibinfo{author}{\bibfnamefont{J.}~\bibnamefont{Heyd}},
  \bibinfo{author}{\bibfnamefont{G.~E.} \bibnamefont{Scuseria}},
  \bibnamefont{and}
  \bibinfo{author}{\bibfnamefont{M.}~\bibnamefont{Ernzerhof}},
  \bibinfo{journal}{J. Chem. Phys.} \textbf{\bibinfo{volume}{{118}}},
  \bibinfo{pages}{8207} (\bibinfo{year}{2003}).

\bibitem[{\citenamefont{Henderson et~al.}(2007)\citenamefont{Henderson,
  Izmaylov, Scuseria, and Savin}}]{HenIzmScuSav-JCP-07}
\bibinfo{author}{\bibfnamefont{T.~M.} \bibnamefont{Henderson}},
  \bibinfo{author}{\bibfnamefont{A.~F.} \bibnamefont{Izmaylov}},
  \bibinfo{author}{\bibfnamefont{G.~E.} \bibnamefont{Scuseria}},
  \bibnamefont{and} \bibinfo{author}{\bibfnamefont{A.}~\bibnamefont{Savin}},
  \bibinfo{journal}{J. Chem. Phys.} \textbf{\bibinfo{volume}{127}},
  \bibinfo{pages}{221103} (\bibinfo{year}{2007}).

\end{thebibliography}
\end{document}